\documentclass[10pt, aps,prx,twocolumn,amsmath,amssymb,floatfix]{revtex4-2}

\usepackage{amsmath,amsfonts,amssymb,mathtools}
\usepackage{bm}
\usepackage{siunitx}
\usepackage[dvipsnames]{xcolor}

\usepackage{graphicx}
\usepackage[dvipsnames]{xcolor}
\usepackage{hyperref}

\newcommand{\figref}[2]{\hyperref[#1]{Fig.~\ref*{#1}#2}}
\newcommand{\secref}[1]{\hyperref[#1]{Sec.~\ref*{#1}}}
\newcommand{\eref}[1]{\hyperref[#1]{Eq.~\ref*{#1}}}
\newcommand{\aref}[1]{\hyperref[#1]{Appendix~\ref*{#1}}}

\newcommand{\tavg}[1]{\left\langle #1 \right\rangle}
\newcommand{\dt}{\Delta t}
\newcommand{\Id}{\bm{\mathcal{I}}}
\newcommand{\sst}{{\begin{subarray}{l} \text{single-}\\ \text{site}\end{subarray}}}

\DeclareMathOperator*{\argmax}{arg\,max}

\begin{document}

\title{Transient dynamics of associative memory models}
\author{David G. Clark}
\thanks{Present address: Kempner Institute for the Study of Natural and Artificial Intelligence, Harvard University, Cambridge, Massachusetts 02138, USA}
\thanks{\href{mailto:dgclark@fas.harvard.edu}{dgclark@fas.harvard.edu}}
\affiliation{Zuckerman Institute, Columbia University, New York, NY 10027, USA}
\affiliation{Kavli Institute for Brain Science, Columbia University, New York, NY 10027, USA}

\date{\today}

\begin{abstract}
Associative memory models such as the Hopfield network and its dense generalizations with higher-order interactions exhibit a ``blackout catastrophe''---a discontinuous transition where stable memory states abruptly vanish when the number of stored patterns exceeds a critical capacity. This transition is often interpreted as rendering networks unusable beyond capacity limits. We argue that this interpretation is largely an artifact of the equilibrium perspective. We derive dynamical mean-field equations for graded-activity dense associative memory models, with the Hopfield model as a special case, using a bipartite cavity approach. We solve the resulting self-consistent equations using an iterative numerical scheme. We show that patterns can be transiently retrieved with high accuracy above capacity despite the absence of stable attractors. This occurs because slow regions persist in the above-capacity energy landscape near stored patterns as lingering traces of the stable basins that existed below capacity. The same transient-retrieval effect occurs in below-capacity networks initialized outside basins of attraction. ``Transient-recovery curves'' provide a concise visual summary of these effects, revealing graceful, non-catastrophic changes in retrieval behavior above capacity and allowing us to compare the behavior across interaction orders. This dynamical perspective reveals energy landscape structure obscured by equilibrium analysis, including slow regions near stored patterns that persist above capacity, and suggests biological neural circuits may exploit transient dynamics for memory retrieval. Furthermore, our approach suggests ways of understanding computational properties of neural circuits without reference to fixed points and yields new theoretical results on generalizations of the Hopfield model.
\end{abstract}

\maketitle

% \tableofcontents

\section{Introduction}
\label{sec:intro}

The Hopfield model is a recurrent neural network with weights constructed through a Hebbian learning rule that can store and retrieve patterns, therefore functioning as a memory device \cite{hopfield1982neural, hopfield1984neurons}. Its dynamics are governed by an energy function, permitting its analysis within equilibrium statistical mechanics, in particular, using methods for disordered systems such as the replica method \cite{amit1985storing, amit1987statistical}.

A key result from this line of work is that the standard Hopfield model can successfully store and retrieve $P = \mathcal{O}(N)$ random patterns as stable fixed points, where $N$ is the number of neurons. Beyond this capacity, interference among patterns encoded in the connectivity destroys these stable states \cite{amit1985storing, amit1987statistical}, a phenomenon known as ``blackout catastrophe'' \cite{tyulmankov2024computational, zenke2024theories}. This represents a discontinuous, first-order phase transition: the overlap between network activity and a target pattern remains high, corresponding to memory retrieval, until a critical capacity of $P \approx 0.14 N$ for binary-spin models. Beyond this capacity, the high-overlap solution vanishes and only the zero-overlap solution remains.

The limited capacity of the Hopfield model has motivated various generalizations. The dense associative memory model, introduced and studied decades ago \cite{abbott1987storage, gardner1987multiconnected, horn1988capacities, chen1986high} and connected to modern deep learning by Krotov and Hopfield \cite{krotov2016dense, krotov2021large}, achieves capacity $P = \mathcal{O}(N^n)$ using $(n+1)$-way neuronal interactions. This represents a qualitative improvement over the Hopfield model's $P = \mathcal{O}(N)$ capacity for $n > 1$, with the Hopfield model corresponding to the special case $n = 1$ with pairwise interactions. These generalized models thus liberate storage capacity from pattern dimensionality, which are constrained to be proportional in the Hopfield case. A caveat to this line of work is that while the model achieves $P = \mathcal{O}(N^n)$ capacity, it can equivalently be formulated as a bipartite system with $P + N$ units using pairwise interactions, thus recovering linear scaling in the total number of units. Rather than appealing to this bipartite formulation, biological implementations of effectively higher-order neuronal interactions have been proposed, but remain speculative \cite{kozachkov2025neuron}. Regardless of how significant one finds the increased capacity, what remains interesting and nontrivial are the \textit{pattern retrieval dynamics}, the focus of this work, that allow these models to recall stored patterns from partial or corrupted inputs within densely packed feature spaces.

Like the Hopfield model, these generalized models possess energy functions governing their dynamics and exhibit discontinuous vanishing of stable retrieval states when capacity is exceeded \cite{abbott1987storage}, though this is less limiting in practice given their increased capacity for $n>1$.

The capacity constraints that seemingly render associative memory models unusable beyond their capacity limits are derived from equilibrium analyses that probe stable memory states, that is, energy landscape local minima, but provide limited insight into the system's behavior during transient evolution. We therefore adopt a dynamical rather than an equilibrium perspective on associative memory models. We study the out-of-equilibrium, transient dynamics of these systems and demonstrate that the blackout catastrophe is not catastrophic when viewed dynamically. In particular, even when stable fixed points no longer exist beyond the critical capacity, memories can still be transiently recalled, often with high accuracy, during evolution from initial conditions. Analysis of the energy function reveals that this transient retrieval reflects the persistence of slow regions in the energy landscape near stored patterns, even above capacity.

Neural circuit dynamics are frequently characterized through fixed points and stability analysis. Indeed, constructing a ``dynamical skeleton'' based on fixed points and transitions between them has been among the most successful approaches for understanding artificial \cite{maheswaranathan2019universality} and biological recurrent neural circuits \cite{vyas2020computation}. However, transient dynamics outside of fixed points are likely crucial for neural computation and require new analysis methods \cite{turner2021charting}. For memory systems, we propose ``transient-recovery curves,'' which characterize memory retrieval performance without requiring stable attractor states. These curves reveal a graceful degradation of retrieval performance as capacity is exceeded, rather than abrupt failure. By varying the number of stored patterns, these curves sweep out a family that can be compared across different interaction orders $n$, permitting analysis of how higher-order interactions shape retrieval dynamics.

To study these transient dynamics in the $N \rightarrow \infty$ limit, we develop a dynamical mean-field theory (DMFT) for graded-activity dense associative memory models storing an appropriately scaled infinite number of patterns. We solve the resulting self-consistent equations using iterative numerical methods similar to those described by \citet{roy2019numerical} for ecological systems. The equilibrium statistical mechanics of graded-activity Hopfield models (the $n=1$ case) was analyzed by \citet{kuhn1991statistical}. While several works from the 1980s and 1990s derived DMFT equations for Hopfield models, they typically studied binary spins in discrete time rather than continuous variables in continuous time and, crucially, could not numerically solve the self-consistent equations. The binary-spin restriction limits relevance to machine learning, which requires differentiability. We review this historical context in detail in the \hyperref[sec:discussion]{Discussion}. The iterative scheme we use here is made feasible by modern computational resources, particularly GPU acceleration.
Our approach allows us to capture the full temporal evolution of these systems, reveals rich transient dynamics that were previously inaccessible to theoretical analysis, and extends all of these analyses to higher-order generalizations of the Hopfield model.

\section{Hopfield and dense associative memory models}
\label{sec:model}

\begin{figure*}
    \centering
    \includegraphics[width=6.2in]{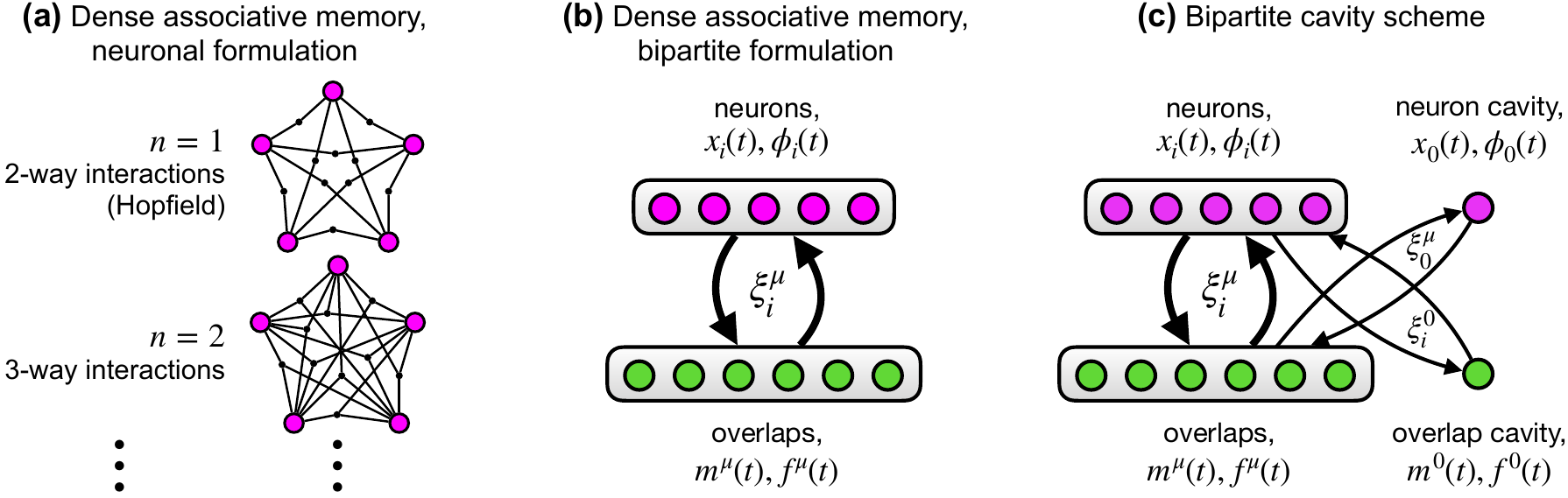}
    \caption{Schematics of the dense associative memory model. \textbf{(a)} Neuronal formulation with higher-order interactions. Nodes represent neurons and black dots indicate connections (tensor elements $T_{i j_1 \cdots j_n}$). \textbf{(b)} Equivalent formulation as a bipartite network with neurons $x_i(t)$ and overlaps $m^\mu(t)$ connected in a bipartite manner through stored patterns $\xi_i^\mu$. \textbf{(c)} Schematic of bipartite cavity scheme used to derive the DMFT.}
    \label{fig:fig_0}
\end{figure*}

We now define the class of models considered in this paper: the dense associative memory model for arbitrary $n$, with the Hopfield model corresponding to $n=1$. This can be done through either a neuronal (\figref{fig:fig_0}{a}) or a bipartite (\figref{fig:fig_0}{b}) formulation, the latter of which is amenable to a cavity analysis (\figref{fig:fig_0}{c}). We then specify a generative process for the stored patterns and initial conditions to enable a large-$N$ analysis. Finally, we distinguish between condensed and uncondensed patterns and show through a signal-to-noise argument that the capacity scales as $P = \mathcal{O}(N^n)$.

\subsection{Neuronal formulation}
\label{subsec:higher_order}

The dense associative memory model generalizes the Hopfield model by introducing higher-order, $(n+1)$-way interactions among neurons (\figref{fig:fig_0}{a}; note that prior works typically define $n$ to be the order of interactions itself, that is, what we call $n+1$). Consider $N$ neurons with preactivations $x_i(t)$ and nonlinearly transformed activations $\phi_i(t) = \phi(x_i(t))$, where $i \in \{1,2,\ldots,N\}$ indexes neurons and $\phi(x)$ is a bounded and monotonic neuronal nonlinearity. The neuronal dynamics are governed by
\begin{widetext}
\begin{equation}
    x_i(t) = (1-\dt)x_i(t-1) 
    + \dt\left[ \frac{g}{\sqrt{\alpha}} \sum_{j_1, j_2, \cdots, j_n} T_{i j_1 j_2 \cdots j_n}  \phi_{j_1}(t-1) \phi_{j_2}(t-1) \cdots \phi_{j_n}(t-1) + I_i(t-1)\right], \label{eq:higher_order_dynamics}
\end{equation}
\end{widetext}
where $t \in \{1, 2, \ldots, T\}$ indexes discrete time steps, $\dt$ is the time step size, and $I_i(t)$ are external inputs that serve as source terms in the mean-field analysis (they can be set to zero when not needed for this purpose). The interaction tensor is constructed from $P$ stored patterns $\xi^\mu_i$, where $\mu \in \{1, 2, \ldots, P\}$ indexes patterns, via
\begin{equation}
    T_{i j_1 j_2 \cdots j_n} = \frac{1}{N^n} \sum_{\mu} \xi^\mu_i \xi^\mu_{j_1} \xi^\mu_{j_2} \cdots \xi^\mu_{j_n}. \label{eq:tensor}
\end{equation}
The memory load parameter, which appears in the dynamics (\eref{eq:higher_order_dynamics}), is defined as
\begin{equation}
    {\alpha = \frac{P}{N^n}.}
\end{equation}

This formulation involves $(n+1)$-way interactions among neurons. For $n=1$, we recover the Hopfield model with pairwise interactions through the matrix
\begin{equation}
    T_{ij} = \frac{1}{N}\sum_{\mu} \xi^\mu_i \xi^\mu_j, \label{eq:hopfield_matrix}
\end{equation}
which can be formed via Hebbian learning with $\xi^\mu_j$ and $\xi^\mu_i$ representing pre- and postsynaptic activity, respectively.

As $\dt \to 0$ while holding the total time $T \dt$ fixed, we obtain the continuous-time limit of the dynamics,
\begin{multline}
    (1 + \partial_t) x_i(t) = \frac{g}{\sqrt{\alpha}} \sum_{j_1, j_2, \cdots, j_n} T_{i j_1 j_2 \cdots j_n} \\
    \times \phi_{j_1}(t) \phi_{j_2}(t) \cdots \phi_{j_n}(t) + I_i(t), \label{eq:continuous_dynamics}
\end{multline}
for which an energy function $\varepsilon[\vec{\phi}]$ can be defined (\secref{subsec:energy}) \cite{krotov2016dense, krotov2021large}. Note that these dynamics are not a gradient flow, since the left-hand side specifies $\partial_t x_i(t)$ while the right-hand side is a gradient with respect to $\phi_i$. Nevertheless, $\varepsilon[\vec{\phi}]$ serves as a Lyapunov function, decreasing monotonically along trajectories due to the monotonicity of $\phi(\cdot)$ (see \secref{subsec:energy} for details).

\subsection{Bipartite formulation}
\label{subsec:bipartite}

This system with higher-order interactions among neurons can be equivalently represented as a bipartite system of neurons and overlaps (\figref{fig:fig_0}{b}). This reformulation proves advantageous for several reasons. First, it provides a pathway for implementing such models using biological neurons and synapses \cite{tyulmankov2021biological}. Second, the bipartite structure lends itself to the cavity method, which we use to derive the DMFT (\figref{fig:fig_0}{c}).

We introduce $P$ overlaps $m^\mu(t)$ defined as
\begin{equation}
    m^\mu(t) = \frac{1}{N}\sum_{i} \xi^\mu_i \phi_i(t) + I^\mu(t), \label{eq:overlap_dynamics}
\end{equation}
where $I^\mu(t)$ are external inputs to the overlaps that serve as source terms in the mean-field analysis. Like the neuronal source terms, they can be set to zero when not needed for this purpose. The overlap $m^\mu(t)$ measures the alignment between the network state at time $t$, $\phi_i(t)$, and the $\mu$-th stored pattern, $\xi^\mu_i$. In analogy to the neuronal nonlinearity, we define nonlinearly transformed overlaps as $f^\mu(t) = f(m^\mu(t))$, where $f(m)$ is a monomial:
\begin{align}
    f(m) &= m^n, \quad n \geq 1. \label{eq:f_nonlin}
\end{align}
The neuronal dynamics of \eref{eq:higher_order_dynamics} can then be written as
\begin{align}
    x_i(t) &= (1-\dt)x_i(t-1) \nonumber \\
    &+ \dt\left[ \frac{g}{\sqrt{\alpha}} \sum_{\mu} \xi^\mu_i f^\mu(t-1)  + I_i(t-1)\right]. \label{eq:neuron_dynamics}
\end{align}
In this bipartite representation, neurons and overlaps interact through couplings given by the stored patterns $\xi^\mu_i$. The Hopfield model corresponds to the case where nonlinearity appears only in the neuronal variables (since, in this case, $f(m) = m$). The dense associative memory model is thus a natural generalization where a pointwise nonlinearity $f(m)$ is also applied to the overlaps.

An interesting extension of these models, which we do not pursue here, is to give the overlaps their own relaxational dynamics rather than the instantaneous equation \eref{eq:overlap_dynamics}. Moreover, the overlaps can have more complex nonlinearities, including nonlinearities that couple different overlaps to each other, while preserving the existence of an energy function. This reveals connections to self-attention mechanisms \cite{krotov2021large}.

\subsection{Pattern and initial-condition statistics}
\label{subsec:statistics}

To analyze the large-$N$ limit, we specify a generative process for the patterns $\xi_i^\mu$, which act as quenched disorder. We adopt the standard assumption of independent and identically distributed pattern components:
\begin{equation}
    \xi^\mu_i \overset{\text{iid}}{\sim} P(\xi),
    \label{eq:pattern_distribution}
\end{equation} 
where $P(\xi)$ is a probability distribution with zero mean and variance $\sigma^2_\xi$.

To study pattern retrieval, we initialize the network with significant overlap with a finite number of patterns of interest, plus random noise. Let $\mu^* \in \{1,2,\ldots,s\}$ index $s$ patterns of interest, where $s$ is finite and typically small (e.g., $s=1$). Neurons are initialized as
\begin{equation}
    x_i(1) = \sum_{\mu^*} a^{\mu^*} \xi^{\mu^*}_i + z_i,
    \label{eq:initial_condition}
\end{equation}
where $a^{\mu^*}$ are coefficients determining the initial overlap with pattern $\mu^*$, and $z_i$ represents random noise independent of the patterns:
\begin{equation}
    z_i \overset{\text{iid}}{\sim} P(z),
    \label{eq:noise_distribution}
\end{equation}
with $P(z)$ having zero mean and variance $\sigma^2_z$.

\subsection{Condensed patterns and capacity}
\label{subsec:capacity}

We now justify the capacity scaling $P = \mathcal{O}(N^n)$ through a signal-to-noise analysis. For this scaling to yield interesting dynamics (e.g., phase transitions at $\mathcal{O}(1)$ values of $\alpha$), the signal from the finite number of patterns of interest and the noise from all other patterns encoded in the weights must compete on equal footing.

A key insight underlying the work of Amit, Gutfreund, and Sompolinsky \cite{amit1985storing, amit1987statistical} is that pattern overlaps have two possible scalings with $N$, which they referred to as condensed and uncondensed patterns. These correspond to signal and noise, respectively:
\begin{equation}
    m^\mu(t) = \begin{cases}
        \mathcal{O}(1) & \text{condensed patterns,} \\
        \mathcal{O}(1/\sqrt{N}) & \text{uncondensed patterns.}
    \end{cases} 
    \label{eq:pattern_scaling}
\end{equation}
The $\mathcal{O}(1)$ overlap signifies that the neuronal state is nontrivially aligned with the corresponding condensed pattern. By contrast, the $\mathcal{O}(1/\sqrt{N})$ overlap for an uncondensed pattern corresponds to the typical inner product, divided by $N$, between two random, independent vectors of dimension $N$. While this scaling is characteristic of independent vectors, uncondensed patterns \textit{do} influence the neuronal state---assuming complete independence between the neuronal state and all patterns would yield incorrect mean-field equations. Condensed and uncondensed patterns are loosely analogous to the rich and lazy regimes of neural-network activity, respectively \cite{van2025coding}. The $s$ patterns used for network initialization (\secref{subsec:statistics}) are the condensed patterns, since they start with $\mathcal{O}(1)$ overlap and maintain this scaling throughout the dynamics. The remaining patterns stay uncondensed, as patterns initialized with $\mathcal{O}(1/\sqrt{N})$ overlap cannot transition to condensed status on $\mathcal{O}(1)$ timescales.

Recall that the neuronal input from stored patterns is given by $\frac{g}{\sqrt{\alpha}} \sum_{\mu} {\xi^\mu_i} f^\mu(t)$ (\eref{eq:neuron_dynamics}). There are $s$ condensed patterns and $P-s$ uncondensed patterns, where $s$ is finite, $P = \alpha N^n$, and $N \rightarrow \infty$. For condensed patterns, we have $f^\mu(t) = [m^\mu(t)]^n = \mathcal{O}(1)$, giving an $\mathcal{O}(1)$ contribution to the neuronal input. Each uncondensed pattern contributes much less individually than condensed patterns, but there are many more of them. Since $m^\mu(t) = \mathcal{O}(1/\sqrt{N})$ for uncondensed patterns, we have $f^\mu(t) = [m^\mu(t)]^n = \mathcal{O}(1/N^{n/2})$. To estimate the total contribution from uncondensed patterns, we treat the quenched disorder $\xi^\mu_i$ and dynamic variables $f^\mu(t)$ as independent. While this approximation would yield incorrect mean-field equations, as mentioned above, it suffices for determining the correct scaling behavior. The input from uncondensed patterns has zero mean, and its magnitude scales as
\begin{equation}
    \underbrace{\frac{g}{\sqrt{\alpha}}}_{\text{prefactor}} \times \underbrace{\sqrt{P}}_{\text{num. terms in sum}} \times \underbrace{\frac{\sigma_\xi}{N^{n/2}}}_{\text{size of each term}} = g\sigma_\xi,\label{eq:noise_scaling}
\end{equation}
where we have used $P = \alpha N^n$. This shows that the noise contribution from uncondensed patterns is also $\mathcal{O}(1)$, confirming that signal and noise compete on equal footing for the chosen scaling $P = \alpha N^n$.

\section{Dynamical Mean-Field Theory (DMFT)}
\label{sec:dmft}

To analyze the transient dynamics of these models in the large-$N$ limit, we develop a DMFT. Unlike traditional equilibrium approaches that focus on fixed points and their stability, DMFT captures the full time evolution of the system, including out-of-equilibrium states where the most interesting memory retrieval properties emerge.

\subsection{Order parameters}
\label{subsec:order_parameters}

The DMFT involves three types of order parameters that characterize macroscopic network activity. The first is the two-time correlation function of neuronal activations,
\begin{align}
        C^\phi(t,t') &= \frac{1}{N}\sum_{i} \phi_i(t) \phi_i(t').
        \label{eq:correlation}
\end{align}
The second is the response function,
\begin{align}
        S^\phi(t,t') &= \frac{1}{N} \sum_{i} \frac{d\phi_i(t)}{d I_i(t')},
        \label{eq:response}
\end{align}
which measures how neuronal activations at time $t$ respond to infinitesimal perturbations of the source term $I_i(t')$ at time $t'$. The third consists of the overlaps with the $s$ condensed patterns used to initialize the dynamics,
\begin{align}
        m^{\mu^*}(t) &= \frac{1}{N} \sum_{i} \xi^{\mu^*}_i \phi_i(t). \label{eq:overlap}
\end{align}
The DMFT consists of self-consistent equations that determine these order parameters in the limit $N \rightarrow \infty$. Finite-size, large-$N$ simulations are expected to match these limiting values up to $\mathcal{O}(1/\sqrt{N})$ fluctuations. 

\subsection{Approaches}

Two main approaches exist for deriving DMFT equations for disordered dynamical systems: path-integral methods and cavity methods. Both approaches address the central challenge that quenched disorder (stored patterns) and dynamic variables (neurons and overlaps) are correlated, making naive disorder averaging incorrect.
In the \hyperref[sec:discussion]{Discussion}, we review path integral methods in detail, as they are the basis of most prior work on the DMFT of Hopfield models. 

In this paper we use the cavity method, which provides a more intuitive approach for handling correlations between quenched disorder and dynamic variables. The basic idea is to remove a dynamic variable from the system, creating the titular cavity, then reintroduce it to analyze its effect on the network perturbatively. This approach is particularly well-suited for bipartite systems \cite{clark2025connectivity, clark2025simplified} like our neuron-overlap formulation. 

The cavity procedure consists of four steps (\figref{fig:fig_0}{c}).
\begin{enumerate}
    \item Begin with an unperturbed system of dynamic variables for a given realization of quenched disorder.
    \item Couple a new ``cavity'' variable to the existing variables via new random couplings. The cavity variable's introduction perturbs the existing variables.
    \item Write the dynamic equation for the cavity variable, where the input it receives from other variables accounts for how those variables are perturbed in response to the cavity variable's introduction. This perturbation generates a self-coupling term in the resulting single-site dynamics.
    \item Average over the quenched disorder to obtain statistics of the quantities appearing in this single-site picture. The cavity construction allows these averages to be computed because, in the expressions of interest, the relevant quenched disorder---the new random couplings between the cavity variable and the original system---is independent of the dynamic variables. This independence holds because the dynamic variables were defined for the original, unperturbed system, before these new couplings were introduced.
\end{enumerate}
The bipartite structure of the system further simplifies this analysis. When introducing a cavity variable (either a neuron or an overlap), we only need to compute its effect on the opposite type of variables (overlaps or neurons, respectively), since only the opposite type provides direct input to the cavity variable. We perform the cavity analysis twice---once with a neuron cavity and once with an overlap cavity---producing two complementary pictures. The self-consistent equations in each picture depend on statistical averages from the other, creating a closed, mutually referential system that determines the order parameters.

The calculation used here is ``zero temperature'' in the sense that the dynamic variables follow deterministic evolution given the quenched disorder. Such zero-temperature cavity methods \cite{mezard2003cavity} have been applied to static problems, including problems with a bipartite structure \cite{ramezanali2015cavity, rocks2022memorizing}. For a cavity calculation of the Hopfield equilibrium properties at finite temperature, see \cite{shamir2000thouless}.

\subsection{Derivation using the bipartite cavity method}

\label{subsec:cavity}
\subsubsection{Neuron cavity}
\label{subsubsec:neuron_cavity}
We first add a ``cavity neuron'' $x_0(t)$ with activation $\phi_0(t)$ to the system. This neuron connects to all existing overlaps through new random couplings $\xi^\mu_0$. The addition of this neuron perturbs the overlaps of uncondensed patterns by
\begin{equation}
    \delta f^\mu(t) = \sum_{t'} \sum_\nu \frac{d f^\mu(t)}{dI^\nu(t')} \frac{1}{N} \xi^\nu_0 \phi_0(t').
    \label{eq:pattern_modification}
\end{equation}
The dynamic equation for the cavity neuron, including feedback in response to its own presence, is
\begin{widetext}
\begin{multline}
    x_0(t) = (1-\dt)x_0(t-1) \\+ \dt\Bigg{[}\underset{\text{from condensed patterns}}{\underbrace{\frac{g}{\sqrt{\alpha}} \sum_{\mu^*} \xi^{\mu^*}_0 f^{\mu^*}(t-1)}}  
    + \underbrace{\frac{g}{\sqrt{\alpha}} \sum_\mu \xi^\mu_0 f^\mu(t-1)}_{\substack{= \eta_0(t-1),\\\text{neuronal cavity field}}} 
    + \sum_{t'}\underbrace{\left[ \frac{g}{\sqrt{\alpha} N} \sum_{\mu,\nu}\xi^\mu_0 \xi^\nu_0 \frac{df^\mu(t-1)}{dI^\nu(t')}  \right]}_{\substack{=F_{00}(t-1,t'),\\\text{neuronal self-coupling kernel}}}~\phi_0(t') + I_0(t-1)\Bigg{]},
    \label{eq:cavity_neuron_dynamics}
\end{multline}
\end{widetext}
where we have separated the contributions from condensed patterns $\mu^*$ and defined the neuronal cavity field and self-coupling kernel. 

The key advantage of the cavity construction is that, as described in step 4 above, it decouples the quenched disorder $\xi^\mu_0$ from the dynamic variables $f^\mu(t)$ and $df^\mu(t)/dI^\nu(t')$, allowing us to evaluate disorder-averaged moments of the cavity field and self-coupling kernel. Here and throughout the derivation, $\tavg{\cdot}$ denotes an average over the quenched disorder, namely, the random patterns $\xi^\mu_i$, $\xi^0_i$, and $\xi^\mu_0$. By the central limit theorem, the neuronal cavity field $\eta_0(t)$ is Gaussian with statistics
\begin{align}
    \tavg{\eta_0(t)} &= \frac{g}{\sqrt{\alpha}} \sum_\mu \underbrace{\tavg{\xi^\mu_0}}_{=0} \tavg{f^\mu(t)} = 0, \label{eq:eta_mean} \\
    \tavg{\eta_0(t)\eta_0(t')} &= \frac{g^2}{\alpha}\sum_{\mu,\nu} \underbrace{\tavg{\xi^\mu_0 \xi^\nu_0}}_{=\delta^{\mu\nu}\sigma^2_\xi} \tavg{f^\mu(t) f^\nu(t')} \nonumber \\
    &= g^2 \sigma^2_\xi N^n \tavg{f^\mu(t) f^\mu(t')}.
    \label{eq:eta_correlation}
\end{align}
Noting that $f^\mu(t) = \mathcal{O}(1/N^{n/2})$, the correlation function $\tavg{\eta_0(t)\eta_0(t')}$ is $\mathcal{O}(1)$, meaning that $\eta_0(t)$ itself is $\mathcal{O}(1)$, as expected from the scaling arguments in \secref{subsec:capacity} (in which the decoupling achieved here using the cavity construction was instead incorrectly assumed). Meanwhile, the self-coupling kernel $F_{00}(t,t')$ is $\mathcal{O}(1)$ and self-averaging with mean
\begin{align}
    F_{00}(t,t')
    &= \frac{g}{\sqrt{\alpha}N} \sum_{\mu,\nu} \underbrace{\tavg{\xi^\mu_0 \xi^\nu_0}}_{=\delta^{\mu\nu}\sigma^2_\xi} \tavg{\frac{df^\mu(t)}{dI^\nu(t')}} \nonumber \\
    &= g\sigma^2_\xi\sqrt{\alpha} N^{n-1} \tavg{\frac{df^\mu(t)}{dI^\mu(t')}}.
    \label{eq:F_mean}
\end{align}

These averages depend on pattern statistics, which we determine through the complementary overlap-cavity analysis:
\begin{align}
    \tavg{\eta_0(t) \eta_0(t')} &= g^2 \sigma_\xi^2 N^n \tavg{f^0(t) f^0(t')} ,\label{eq:eta_nrn} \\
    {F}_{00}(t,t') &= g \sigma_\xi^2\sqrt{\alpha} N^{n-1} \tavg{\frac{df^0(t)}{d{I}^0(t')}}. \label{eq:F_nrn}
\end{align}
To complete the analysis, we now derive the pattern overlap-cavity equations that will allow us to compute these averages.

\subsubsection{Overlap cavity}
\label{subsubsec:pattern_cavity}
We add an overlap $m^0(t)$, for an uncondensed pattern, connected to all neurons through new random couplings $\xi^0_i$. The perturbation to neurons due to this cavity overlap is
\begin{equation}
    \delta \phi_i(t) = \sum_{t'} \sum_j \frac{d\phi_i(t)}{dI_j(t')} \frac{g}{\sqrt{\alpha}} \xi^{0}_j f^{0}(t').
    \label{eq:neuron_modification}
\end{equation}
The dynamic equation for the cavity overlap is then
\begin{widetext}
\begin{align}
    m^{0}(t) &= \underbrace{\frac{1}{N} \sum_i \xi^{0}_i \phi_i(t)}_{=\eta^{0}(t), \text{ overlap cavity field}} + \sum_{t'} \underbrace{\left[ \frac{g}{\sqrt{\alpha}N} \sum_{i,j} \xi^{0}_i \xi^{0}_j \frac{d\phi_i(t)}{dI_j(t')}\right]}_{\substack{=F^{00}(t,t'), \text{ overlap self-coupling kernel}}} f^{0}(t') + I^{0}(t),
    \label{eq:cavity_pattern_dynamics}
\end{align}
\end{widetext}
where we have defined the overlap cavity field and self-coupling kernel. As with the neuronal cavity, we use the independence of quenched disorder and dynamic variables to evaluate disorder averages. The cavity field $\eta^0(t)$ is Gaussian with statistics
\begin{align}
    \tavg{\eta^{0}(t)} &= \frac{1}{N} \sum_i \underbrace{\tavg{\xi^{0}_i}}_{=0} \tavg{\phi_i(t)} = 0, \label{eq:eta_nrn_mean} \\
    \tavg{\eta^{0}(t) \eta^{0}(t')} &= \frac{1}{N^2} \sum_{i,j} \underbrace{\tavg{\xi^{0}_i \xi^{0}_j}}_{=\delta_{ij}\sigma^2_\xi} \tavg{\phi_i(t) \phi_j(t')} \nonumber \\
    &= \frac{\sigma^2_\xi}{N} C^\phi(t,t').
    \label{eq:eta_nrn_correlation}
\end{align}
The correlation function $\tavg{\eta^{0}(t) \eta^{0}(t')}$ is $\mathcal{O}(1/N)$, implying that $\eta^0(t) = \mathcal{O}(1/\sqrt{N})$, consistent with the overlap itself being $\mathcal{O}(1/\sqrt{N})$ as expected for an uncondensed pattern. Meanwhile, the self-coupling kernel is $\mathcal{O}(1)$ and self-averaging with mean
\begin{align}
    F^{00}(t,t') &= \frac{g}{\sqrt{\alpha}N} \sum_{i,j} \underbrace{\tavg{\xi^{0}_i \xi^{0}_j}}_{=\delta_{ij}\sigma^2_\xi} \tavg{\frac{d\phi_i(t)}{dI_j(t')}} \nonumber \\
    &= \frac{g\sigma^2_\xi}{\sqrt{\alpha}} S^\phi(t,t').
    \label{eq:F_nrn_mean}
\end{align}
Thus the overlap cavity picture's cavity field and self-coupling kernel depend on the neuronal order parameters $C^\phi(t,t')$ and $S^\phi(t,t')$, which can be determined within the neuronal cavity picture. This creates mutually referential cavity pictures that together determine the order parameters.

\subsubsection{Evaluating correlation and response functions}
\label{subsec:response_correlation}
Since we aim to determine the neuronal order parameters $C^\phi(t,t')$, $S^\phi(t,t')$, and $m^{\mu^*}(t)$, it is useful to close the mean-field equations in neuronal quantities. To do this, we must evaluate the neuronal cavity-field correlation (\eref{eq:eta_nrn}) and self-coupling kernel (\eref{eq:F_nrn}). To deal with temporal indices, it is helpful to use the following vector and matrix notation:
\begin{itemize}
    \item For a time-dependent scalar quantity $q(t)$ with $t \in \{1, 2, \ldots, T\}$, we define the corresponding $T$-dimensional vector $\bm{q}$ with components $[\bm{q}]_t = q(t)$.
    \item For a two-time function $M(t,t')$ with $t,t' \in \{1, 2, \ldots, T\}$, we define the corresponding $T \times T$ matrix $\bm{M}$ with elements $[\bm{M}]_{t,t'} = M(t,t')$.
    \item For a two-time derivative $d \psi(t) / dI(t')$, we define the corresponding $T \times T$ matrix $d\bm{\psi} / d\bm{I}^T$ with elements $[d\bm{\psi} / d\bm{I}^T]_{t,t'} = d \psi(t) / dI(t')$.
\end{itemize}
In this notation, \eref{eq:eta_nrn} and \eref{eq:F_nrn} for the neuronal cavity-field correlation and self-coupling kernel, respectively, are
\begin{align}
    \tavg{\bm{\eta}_0 \bm{\eta}_0^T} &= g^2 \sigma_\xi^2 N^n \tavg{f(\bm{m}^0) f(\bm{m}^0)^T}, \label{eq:eta_nrn_corr_matrix} \\
    \bm{F}_{00} &= g \sigma_\xi^2\sqrt{\alpha} N^{n-1} \tavg{\frac{df(\bm{m}^0)}{d(\bm{I}^0)^T}}, \label{eq:F_nrn_matrix}
\end{align}
where $f(\bm{m}^0)$ applies the nonlinearity elementwise to the vector $\bm{m}^0$. We need to evaluate these expressions to leading order, namely $\mathcal{O}(1)$.
From the overlap cavity picture, $\bm{m}^0$ obeys, in matrix notation,
\begin{align}
\bm{m}^0 &= \bm{\eta}^0 + \frac{g \sigma_\xi^2}{\sqrt{\alpha}} \bm{S}^\phi f(\bm{m}^0) + \bm{I}^0 \label{eq:m_matrix} \\
\text{where} \quad \tavg{\bm{\eta}^0} &= 0, \quad
\tavg{\bm{\eta}^0 (\bm{\eta}^0)^T} = \frac{\sigma_\xi^2}{N}\bm{C}^\phi .\label{eq:eta_nrn_matrix}
\end{align}
Here, $\bm{\eta}^0$ is Gaussian and $\bm{m}^0$ is determined by solving the nonlinear equation \eref{eq:m_matrix}.

At this point, the analysis diverges between the Hopfield ($n=1$) and higher-order models ($n>1$). In the $n=1$ case, the overlap equation \eref{eq:m_matrix} is linear, which simplifies the calculations. For $n>1$, \eref{eq:m_matrix} is nonlinear, but the nonlinear self-interaction is smaller than $\bm{\eta}^0$ by a factor of $1/N^{(n-1)/2}$, allowing for a perturbative treatment.

Before handling each case, we derive a general expression for the response-function term $\tavg{d\bm{m}^0 / d(\bm{I}^0)^T}$ that applies to both cases. Differentiating both sides of \eref{eq:m_matrix} with respect to $\bm{I}^0$ and solving for $d\bm{m}^0/d(\bm{I}^0)^T$ gives
\begin{equation}
    \frac{d\bm{m}^{0}}{d(\bm{I}^{0})^T} = \left( 
    \Id - \frac{g\sigma^2_\xi}{\sqrt{\alpha}} \bm{S}^\phi \mathcal{D}[f'(\bm{m}^{0})]
    \right)^{-1},
    \label{eq:m_response}
\end{equation}
where $\mathcal{D}[\cdot]$ denotes a diagonal matrix with the argument vector on the diagonal when applied to a vector, or zeros out the off-diagonal elements when applied to a matrix; $\Id$ is the identity matrix; and we set $\bm{I}^0 = \bm{0}$. This allows us to express \eref{eq:F_nrn_matrix} as
\begin{multline}
    \bm{F}_{00} = g\sigma^2_\xi\sqrt{\alpha}N^{n-1} \\
    \times \tavg{\mathcal{D}[f'(\bm{m}^{0})]\left( 
    \Id - \frac{g\sigma^2_\xi}{\sqrt{\alpha}} \bm{S}^\phi \mathcal{D}[f'(\bm{m}^{0})]
    \right)^{-1}}.
    \label{eq:F_expanded}
\end{multline}

\textbf{Hopfield model ($n=1$).}
For the Hopfield model, due to the linearity of the overlap dynamics, we have
\begin{align}
    \bm{m}^0 &= \frac{d\bm{m}^0}{d(\bm{I}^0)^T} \bm{\eta}^0, \\
    \text{where} \quad \frac{d\bm{m}^{0}}{d(\bm{I}^{0})^T} &= \left( 
    \Id - \frac{g\sigma^2_\xi}{\sqrt{\alpha}} \bm{S}^\phi 
    \right)^{-1}.
    \label{eq:m_linear_response}
\end{align}
Thus, \eref{eq:eta_nrn_corr_matrix} becomes
\begin{align}
    \tavg{\bm{\eta}_0\bm{\eta}_0^T} &= g^2 \sigma_\xi^2 N \tavg{\bm{m}^0 (\bm{m}^0)^T} \nonumber \\
    &= g^2 \sigma_\xi^2 N \tavg{\frac{d\bm{m}^0}{d(\bm{I}^0)^T} \bm{\eta}^0 (\bm{\eta}^0)^T \left(\frac{d\bm{m}^0}{d(\bm{I}^0)^T}\right)^T} \nonumber \\
    &= g^2 \sigma_\xi^4 \left(\Id - \frac{g\sigma^2_\xi}{\sqrt{\alpha}} \bm{S}^\phi\right)^{-1} \bm{C}^\phi \left(\Id - \frac{g\sigma^2_\xi}{\sqrt{\alpha}} \bm{S}^\phi\right)^{-T}.
    \label{eq:eta_hopfield}
\end{align}
Meanwhile, since $f(m) = m$, \eref{eq:F_expanded} simplifies to
\begin{equation}
    \bm{F}_{00} = g\sigma_\xi^2\sqrt{\alpha} \left(\Id - \frac{g\sigma^2_\xi}{\sqrt{\alpha}} \bm{S}^\phi\right)^{-1}.
    \label{eq:F_hopfield}
\end{equation}

\textbf{Higher-order models ($n>1$)}.
For $n > 1$, the pattern equation \eref{eq:m_matrix} is nonlinear, but, as mentioned above, the nonlinear self-coupling term $f(\bm{m}^0)$ is smaller than the Gaussian input $\bm{\eta}^0$ by a factor of $1/N^{(n-1)/2}$, permitting a perturbative treatment. For \eref{eq:eta_nrn_corr_matrix}, to leading order, we can replace $\bm{m}^0$ with $\bm{\eta}^0$:
\begin{align}
    \tavg{\bm{\eta}_0 \bm{\eta}_0^T} &= g^2 \sigma_\xi^2 N^n \tavg{f(\bm{\eta}^0) f(\bm{\eta}^0)^T}  \nonumber \\
    &= g^2 \sigma_\xi^{2(n+1)} \bm{P}_{n,n},
    \label{eq:eta_P_matrix}
\end{align}
where we define the matrices
\begin{equation}
    \bm{P}_{n,n'} = \tavg{\bm{u}^n (\bm{u}^{n'})^T}_{\bm{u} \sim \mathcal{N}(0, \bm{C}^\phi)}, 
    \label{eq:P_matrix}
\end{equation}
with powers applied elementwise.

For the response function, noting that $f'(\bm{m}^{0}) = \mathcal{O}(1/N^{(n-1)/2})$, we expand the matrix inverse in \eref{eq:F_expanded}:
\begin{widetext}
\begin{align}
    \bm{F}_{00} &= g\sigma_\xi^2 \sqrt{\alpha} N^{n-1} \tavg{\mathcal{D}[f'(\bm{m}^0)]\left(\Id + \frac{g\sigma_\xi^2}{\sqrt{\alpha}}\bm{S}^\phi \mathcal{D}[f'(\bm{m}^0)] + \cdots \right)} \nonumber \\
    &= g \sigma_\xi^2 \sqrt{\alpha} N^{n-1} \tavg{\mathcal{D}[f'(\bm{m}^0)]} +
    g^2 \sigma_\xi^4  \bm{S}^\phi \circ \tavg{N^{n-1} f'(\bm{m}^0) f'(\bm{m}^0)^T}  + \cdots
    \label{eq:F_higher_expanded}
\end{align}
\end{widetext}
where $\circ$ denotes the Hadamard product. 

For odd $n$, $f'(m) = n m^{n-1}$ is an even function, so $\tavg{f'(\bm{m}^{0})} \neq 0$ and is $\mathcal{O}(1/N^{(n-1)/2})$. The first term in the expansion (\eref{eq:F_higher_expanded}) then scales as $\mathcal{O}(N^{(n-1)/2})$, which diverges. This divergence could presumably be removed through exclusion of self-interactions in \eref{eq:higher_order_dynamics}. We leave this to future work and restrict to even $n$, for which $f'$ is odd and the problematic term vanishes. The different scaling behaviors of the neuronal input for even and odd $n$ can be illustrated through a simple numerical experiment (\aref{sec:odd_n}, \figref{fig:odd_n}{}).

For even $n$, we use \eref{eq:m_matrix} to iteratively express $f'(\bm{m}^0)$ in terms of $\bm{\eta}^0$:
\begin{equation}
    f'(\bm{m}^0) = f'(\bm{\eta}^0) + \frac{g\sigma_\xi^2}{\sqrt{\alpha}} \mathcal{D}[f''(\bm{\eta}^0)] \bm{S}^\phi f(\bm{\eta}^0) + \cdots,
    \label{eq:f_prime_expansion}
\end{equation}
whose $k$-th term is $\mathcal{O}(1/N^{k(n-1)/2})$. Substitution into the first term of the $\bm{F}_{00}$ expansion \eref{eq:F_higher_expanded} gives
\begin{multline}
    g\sigma_\xi^2 \sqrt{\alpha} N^{n-1} \tavg{\mathcal{D}[f'(\bm{m}^0)]} 
    = g\sigma_\xi^2 \sqrt{\alpha} N^{n-1} \tavg{\mathcal{D}[f'(\bm{\eta}^0)]}  \\
    + g^2 \sigma_\xi^4 N^{n-1} \mathcal{D}[\bm{S}^\phi \tavg{f(\bm{\eta}^0) f''(\bm{\eta}^0)^T}] + \cdots .
    \label{eq:F_first_term}
\end{multline}
Since $n$ is even, $f'(m)$ is an odd function, so $\tavg{f'(\bm{\eta}^0)} = 0$. For the second term in the expansion \eref{eq:F_higher_expanded}, we need $\tavg{N^{n-1} f'(\bm{m}^0) f'(\bm{m}^0)^T}$. To leading order, we can replace $f'(\bm{m}^0)$ with $f'(\bm{\eta}^0)$. Combining these results gives
\begin{multline}
    \bm{F}_{00} = g^2 \sigma_\xi^4 \Bigg[ \mathcal{D}\left[
    \bm{S}^{\phi} \tavg{N^{n-1} f(\bm{\eta}^0) f''(\bm{\eta}^0)^T}\right] \\ + \bm{S}^\phi \circ \tavg{N^{n-1} f'(\bm{\eta}^0) f'(\bm{\eta}^0)^T}
    \Bigg],
    \label{eq:F_higher_final}
\end{multline}
where $\mathcal{D}[\cdot]$ now extracts the diagonal elements.
This can be written
\begin{multline}
    \bm{F}_{00} = g^2 \sigma_\xi^{2(n+1)} \Bigg[ n(n-1)\mathcal{D}\left[
    \bm{S}^{\phi} \bm{P}_{n,n-2}\right] \\
    +
    n^2 \bm{S}^\phi \circ \bm{P}_{n-1,n-1}
    \Bigg].
    \label{eq:F_P_matrix}
\end{multline}

\subsubsection{Final self-consistent system}
\label{subsec:final_system}
In the single-site picture, $x_0(t)$ evolves from an initial condition, driven by a cavity field and self-coupling. The statistics of this single-site process determine the self-coupling kernel and cavity-field correlation function.
\begin{widetext}
\begin{align}
    x_0(1) &= \sum_{\mu^*=1} a^{\mu^*} \xi^{\mu^*}_0 + z_0, \label{eq:x0_init} \\
    x_0(t) &= (1-\dt)x_0(t-1) + \dt\Bigg[\frac{g}{\sqrt{\alpha}}\sum_{\mu^*=1} \xi^{\mu^*}_0 \left(m^{\mu^*}(t-1)\right)^n 
    + \eta_0(t-1) + \sum_{t'=1}^{t-1} F_{00}(t-1,t') \phi_0(t') \Bigg],
    \label{eq:x0_dynamics} \\
    \bm{C}^{\eta_0} &= \begin{cases}
        g^2 \sigma_\xi^4 \left(\Id - \frac{g\sigma^2_\xi}{\sqrt{\alpha}} \bm{S}^\phi\right)^{-1} \bm{C}^\phi \left(\Id - \frac{g\sigma^2_\xi}{\sqrt{\alpha}} \bm{S}^\phi\right)^{-T} & n=1 \\
         g^2 \sigma_\xi^{2(n+1)} \bm{P}_{n,n} & n > 1 \text{, even}
    \end{cases}
    \label{eq:C_eta0_final}  \\
    \bm{F}_{00} &= \begin{cases}
        g\sigma^2_\xi\sqrt{\alpha} \left(\Id - \frac{g\sigma^2_\xi}{\sqrt{\alpha}} \bm{S}^\phi\right)^{-1} & n=1 \\
         g^2 \sigma_\xi^{2(n+1)} \left[ n(n-1)\mathcal{D}\left[
    \bm{S}^{\phi} \bm{P}_{n,n-2}\right] + n^2\bm{S}^\phi \circ \bm{P}_{n-1,n-1}
    \right] & n > 1 \text{, even}
    \end{cases}  \label{eq:F00_final}
\end{align}
\end{widetext}
The DMFT is closed by the following self-consistency conditions, with the single-site average $\tavg{\cdot}_\sst$ defined in \eref{eq:single_site_average}:
\begin{align}
    C^\phi(t,t') &= \tavg{\phi_0(t) \phi_0(t')}_\sst, \label{eq:C_phi_self}
\end{align}
\begin{align}
    S^\phi(t,t') &= \tavg{\frac{d\phi_0(t)}{d I_0(t')}}_\sst, \label{eq:S_phi_self} \\
    m^{\mu^*}(t) &= \tavg{\xi^{\mu^*}_0 \phi_0(t)}_\sst, \label{eq:m_mu_self}
\end{align}
where the single-site average $\tavg{\cdot}_\sst$ denotes averaging over the Gaussian noise realization, condensed pattern components, and initialization noise:
\begin{equation}
\tavg{\cdots}_\sst
    = \tavg{\cdots}_{\begin{subarray}{l}
        {\bm{\eta}}_0 \sim \mathcal{N}(\bm{0}, \bm{C}^{\eta_0} )\\
        \xi^{\mu^*}_0 \overset{\text{iid}}{\sim} P(\xi) \text{ for } \mu^* = 1,\ldots,s\\
        z_0 \sim P(z)
    \end{subarray}} .
    \label{eq:single_site_average}
\end{equation}
These equations form a closed system that can be solved numerically to obtain the dynamical behavior of the model in the $N\rightarrow \infty$ limit.

\subsection{Energy function}
\label{subsec:energy}

The models we study possess energy functions that govern their dynamics \cite{krotov2016dense, krotov2021large}. For a configuration of neuronal activations $\vec{\phi} = \{\phi_i\}_{i=1}^N$, the $\mathcal{O}(1)$ energy is
\begin{equation}
    \varepsilon[\vec{\phi}] = -\frac{g }{(n+1) \sqrt{\alpha}}\sum_{\mu} \left( m^\mu \right)^{n+1} + \frac{1}{N}\sum_{i} \mathcal{F}(\phi_i),
\end{equation}
where $\mathcal{F}(\phi)$ satisfies $\mathcal{F}'(\phi) = \phi^{-1}(\phi)$ (\aref{app:comp_F}). The continuous-time limit of the dynamics obeys
\begin{equation}
    \partial_t x_i(t) = -N \partial_{\phi_i} \varepsilon[\vec{\phi}(t)].
\end{equation}
Note that these are not gradient dynamics because the left-hand side specifies the time derivative of $x_i$ while the right-hand side is a gradient with respect to $\phi_i$. Nevertheless, $\varepsilon[\vec{\phi}]$ acts as a Lyapunov function since the energy decreases monotonically:
\begin{align}
    \partial_t \varepsilon[\vec{\phi}] &= \sum_{i} \partial_{\phi_i} \varepsilon[\vec{\phi}] \phi'_i(t) \partial_t x_i(t) \nonumber \\
    &= -\frac{1}{N} \sum_{i} \phi'_i(t) \left( \partial_{t} x_i(t) \right)^2 \leq 0,
\end{align}
where the inequality follows because the nonlinearity is monotonic, $\phi'(x) > 0$. In the DMFT framework, we can express the energy in terms of order parameters as
\begin{widetext}
\begin{align}
    {\varepsilon}(t) &= 
    -\left\{\begin{array}{ll}
    \frac{\sqrt{\alpha}}{2g}\tavg{\eta_0(t) \eta_0(t)}_\sst & n=1 \\
    \tavg{\eta_0(t) \phi_0(t)}_\sst & n>1,\text{ even}
    \end{array}\right\} - \frac{g }{(n+1)\sqrt{\alpha}}\sum_{\mu^*=1} \big({m}^{\mu^*}(t)\big)^{n+1} + \tavg{\mathcal{F}({\phi}_0(t))}_{\sst}.
\end{align}
\end{widetext}
We found this expression to be the most numerically stable of several equivalent formulations.

\subsection{Numerical solution of the DMFT}
\label{sec:numerical}

We solve the self-consistent DMFT equations using an iterative procedure that samples trajectories and updates order parameters. A very closely related approach for an ecological system is detailed in \cite{roy2019numerical}. The steps are as follows. 
\begin{enumerate}
    \item Initialize order parameters $\bm{C}^\phi$, $\bm{S}^\phi$, and $\bm{m}^{\mu^*}$ for $\mu^* \in \{1,2,\ldots,s\}$.
    \item Sample $M$ noise trajectories ${\bm{\eta}}_{m}$, $m\in\{1,2,\ldots,M\}$, through Cholesky decomposition of $\bm{C}^{\eta_0}$.
    \item Sample $M$ sets of $s$ condensed patterns $\xi^{\mu^*}_{m}$.
    \item Forward integrate the $M$ trajectories using the single-site dynamics to get $\bm{x}_m$ and $\bm{\phi}_m = \phi(\bm{x}_m)$, yielding updated correlation function $\bm{C}^\phi$ and overlaps $\bm{m}^{\mu^*}$ in the straightforward way (\eref{eq:C_update} and \eref{eq:m_update} below).
    \item Compute an updated response function $\bm{S}^\phi$ (described in \secref{subsec:response_computation} below).
    \item Update order parameters with memory factor $\gamma \in [0,1]$:
    \begin{align}
        \bm{C}^\phi_\text{new} &= (1-\gamma)\times \bm{C}^\phi_\text{old} + \gamma \times \frac{1}{M}\sum_{m} \bm{\phi}_m \bm{\phi}_m^T , \label{eq:C_update} \\
        \bm{m}^{\mu^*}_\text{new} &= (1-\gamma)\times \bm{m}^{\mu^*}_\text{old} + \gamma \times \frac{1}{M}\sum_{m} \xi^{\mu^*}_{m}\bm{\phi}_m  \label{eq:m_update}, \\
        \bm{S}^\phi_\text{new} &= (1-\gamma)\times \bm{S}^\phi_\text{old} + \gamma \times \bm{S}^\phi_\text{tot} .\label{eq:S_update}
    \end{align}
    \item Repeat steps 2-6 until convergence of order parameters
\end{enumerate}

\subsubsection{Response function (\texorpdfstring{$S^\phi_{\text{tot}}$}{S phi tot}) computation}
\label{subsec:response_computation}
For each trajectory $m$, the response function is computed through forward integration in $t$ for each fixed $s$. The computation begins at $t=s$ and proceeds forward in time:
\begin{multline}
    S^x_m(t,s) = (1-\dt)S^x_m(t-1,s) \\
    + \dt \Bigg[ \sum_{t'=s}^{t-1}F_{00}(t-1,t') \: \phi'_m(t') \:S^x_m(t',s) + \delta_{t-1,s}\Bigg],
    \label{eq:S_x_forward}
\end{multline}
subject to initial conditions
\begin{equation}
    S^x_m(s,s) = 0.
    \label{eq:S_x_init}
\end{equation}
The equation is applied sequentially for $t = s+1, s+2, \ldots, T$ to build up the full response function. For each trajectory, the activation response function is
\begin{equation}
    S^\phi_m(t,s) = \phi'_m(t) \: S^x_m(t,s).
    \label{eq:S_phi_from_S_x}
\end{equation}
The final response function is obtained by averaging over all trajectories:
\begin{equation}
    S^\phi_{\text{tot}}(t,s) = \frac{1}{M} \sum_{m} S^\phi_m(t,s).
    \label{eq:S_phi_average}
\end{equation}
We implement the numerical solution on a GPU using PyTorch, enabling efficient computation with large sample sizes $M$. The primary computational loops are the $T$ time steps for trajectory evolution and the $T(T-1)/2$ time steps for response function integration. The $\mathcal{O}(T^2)$ response function computation dominates the runtime in practice.

\section{Results}
\label{sec:results}

\begin{figure*}
    \centering
    \includegraphics[width=7in]{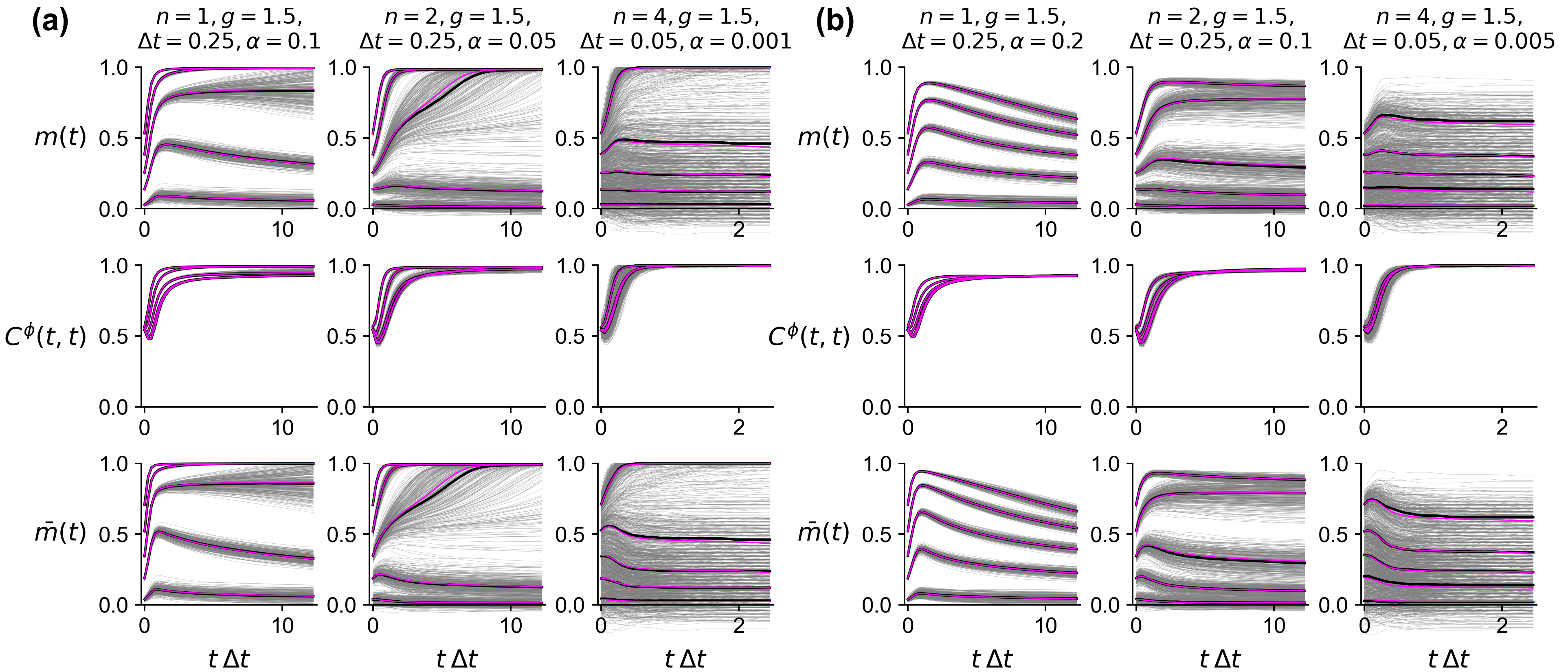}
    \caption{Dynamical evolution of order parameters for Hopfield ($n=1$) and dense associative memory models ($n=2,4$). \textbf{(a)} Below-capacity dynamics with $\alpha = 0.10, 0.05, 0.001$ for $n=1,2,4$, respectively. \textbf{(b)} Above-capacity dynamics with $\alpha = 0.20, 0.10, 0.005$ for $n=1,2,4$, respectively. In \textbf{(a)} and \textbf{(b)}, we show (top) raw overlap $m(t)$, (middle) equal-time correlation $C^\phi(t,t)$, and (bottom) normalized overlap $\bar{m}(t) = m(t)/(\sigma_\xi\sqrt{C^\phi(t,t)})$. {\color{gray}\textbf{Gray}} traces show individual finite-size simulations with $N = 20000, 2000, 200$ for $n = 1, 2, 4$, respectively; \textbf{black} lines show simulation medians; and {\color{magenta}\textbf{magenta}} lines show DMFT predictions.}
    \label{fig:ops_vs_time}
\end{figure*}

\subsection{Simulation setup}

We restrict our analysis to interaction orders $n = 1$, $2$, and $4$. Validation with finite-size simulations requires large $N$ to reduce fluctuations that scale as $\mathcal{O}(1/\sqrt{N})$, but the scaling $P = \mathcal{O}(N^n)$ makes the number of patterns prohibitively large for $n > 4$. We exclude $n = 3$ due to the divergent behavior identified in \secref{sec:dmft}.

All simulations and DMFT solutions focus on retrieval of a single condensed pattern, so we drop the $\mu^*$ superscript and denote the overlap as $m(t)$. We initialize the system according to \eref{eq:initial_condition} with $a = \bar{a} g$, where $\bar{a} \in [0,1]$ controls the initial alignment with the pattern of interest. The initialization noise level is set to $\sigma_z = \sqrt{g^2 - a^2}$ (see \eref{eq:noise_distribution}) so that the variance of the initial condition remains constant as we vary $\bar{a}$. By sweeping $\bar{a}$ from 0 to 1, we explore a range of initial overlaps with the stored pattern.

For simulations, we use system sizes $N = 20000$, $2000$, and $200$ for $n = 1$, $2$, and $4$, respectively, with time steps $\dt = 0.25$ for $n = 1, 2$ and $\dt = 0.05$ for $n = 4$ to ensure numerical stability. We examine the effects of $g$ and $\dt$ in the Hopfield model in \aref{sec:gain}.

\subsection{Equilibrium analysis}

We first confirm that our model exhibits the equilibrium ``blackout catastrophe''---the discontinuous phase transition where high-overlap solutions vanish above a critical capacity. We derive a mean-field theory for fixed-point solutions by removing the time dependencies from our DMFT equations. For the gain parameter $g = 1.5$ used throughout our analyses, we obtain critical capacities $\alpha_c \approx 0.13$, $0.080$, and $0.0011$ for $n = 1$, $2$, and $4$, respectively. Details of this fixed-point analysis are provided in \aref{app:equilibrium}.

\subsection{Retrieval dynamics and key phenomena}

The DMFT solutions show two types of retrieval behaviors.

\textbf{Stable retrieval} occurs when the overlap converges to a value independent of local variations in initial conditions. The convergence is rapid and the asymptotic overlap is close to unity, as expected from prior equilibrium analyses \cite{amit1985storing, amit1987statistical, kuhn1991statistical}.

\textbf{Transient retrieval} occurs when the overlap initially increases but then decreases, failing to reach a stable state. This occurs either when networks exceed capacity or when they are initialized outside basins of attraction. The initial increase is fast, while the eventual decay is much slower, and the late-time overlap value appears to remain nonzero, consistent with the ``remnant overlap'' found in simulations of binary-spin Hopfield models \cite{amit1987statistical}. We cannot determine exact asymptotic values of this remnant overlap because the $\mathcal{O}(T^2)$ time complexity of our numerical solver limits the accessible time horizon.

Unlike in networks of binary spins, the time-dependent variance $C^\phi(t,t)$ is non-trivial, and changes in the overlap $m(t)$ reflect both changes in alignment with the pattern and changes in the variance of the neuronal state. We therefore focus on the normalized overlap,
\begin{equation}
    \bar{m}(t) = \frac{m(t)}{\sigma_\xi\sqrt{C^\phi(t,t)}},
\end{equation}
which lies in $[-1,1]$. We define $\bar{m}_{\text{init}} = \bar{m}(1)$ as the initial normalized overlap.

\subsection{Validation and main results}

We first verify that DMFT solutions match finite-size simulations. \figref{fig:ops_vs_time}{} demonstrates excellent agreement between theory and simulations for the overlap $m(t)$, equal-time correlation function $C^\phi(t,t)$, and normalized overlap $\bar{m}(t)$. 

The normalized overlap exhibits two regimes.

\textbf{Below capacity ($\alpha < \alpha_c$):} When $\bar{m}_{\text{init}}$ is sufficiently large, $\bar{m}(t)$ converges to a stable retrieval state with rapid convergence and final values close to unity. For smaller $\bar{m}_{\text{init}}$, $\bar{m}(t)$ transiently increases and then decays slowly. Thus, even when initialized outside the basin of attraction, patterns can still be transiently recalled.

\textbf{Above capacity ($\alpha > \alpha_c$):} No initial overlap $\bar{m}_{\text{init}}$ is sufficiently large to elicit stable retrieval. Instead, we observe transient retrieval for all $\bar{m}_{\text{init}}$. However, contrary to the equilibrium picture predicting abrupt breakdown, maximum normalized overlaps achieved can be quite high. Since the transient increase is fast and subsequent decay much slower, this behavior can resemble stable retrieval from the below-capacity regime, with the key difference being eventual slow decay rather than convergence to a stable state.

\subsection{Transient-recovery curves}
\label{subsec:transient_recovery}

\begin{figure*}
    \centering
    \includegraphics[width=7in]{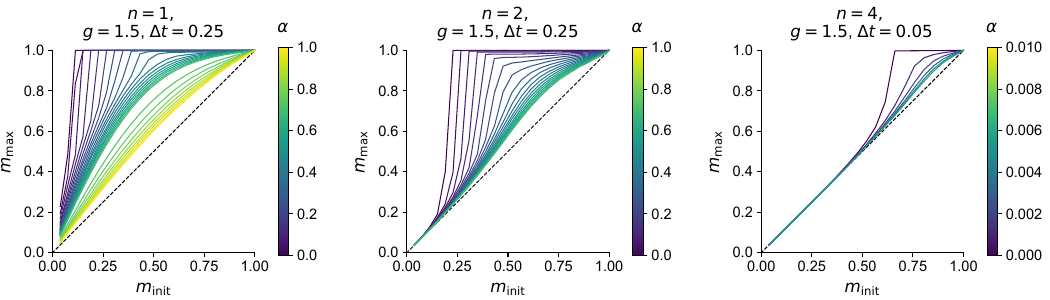}
    \caption{Transient-recovery curves for Hopfield model ($n=1$) and dense associative memory models ($n=2,4$). Each curve plots the maximum normalized overlap $\bar{m}_{\text{max}}$ achieved during dynamical evolution versus the initial normalized overlap $\bar{m}_{\text{init}}$. Different curves within each panel correspond to different memory loads $\alpha = P/N^n$. The diagonal line is the trivial lower bound where maximum overlap equals initial overlap.}
\label{fig:trans_recov_curves}
\end{figure*}

To quantify memory retrieval performance beyond stable attractors, we introduce \textit{transient-recovery curves}. These curves characterize a network's ability to recall stored patterns by plotting the maximum normalized overlap achieved during the entire dynamical evolution,
\begin{equation}
    \bar{m}_{\text{max}} = \max_{t} \bar{m}(t),
\end{equation}
as a function of the initial normalized overlap $\bar{m}_{\text{init}}$. Each curve captures the best retrieval performance accessible through transient dynamics, regardless of whether this optimal retrieval occurs at a stable fixed point or during a transient. \figref{fig:trans_recov_curves}{} shows these transient-recovery curves for the Hopfield model and dense associative memory models with $n = 2$ and $n = 4$, where each curve corresponds to a different memory load $\alpha$.

All curves lie above the diagonal since $\bar{m}_{\text{max}} \geq \bar{m}_{\text{init}}$. When $\alpha$ is sufficiently small and $\bar{m}_{\text{init}}$ is sufficiently large, the network enters a stable retrieval state. This is reflected by the curve becoming flat. That is, $\bar{m}_{\text{max}}$ becomes insensitive to local changes in $\bar{m}_{\text{init}}$. Below capacity but outside the basin of attraction, $\bar{m}_{\text{max}}$ increases smoothly with $\bar{m}_{\text{init}}$ until entering the basin, reflecting transient retrieval. Above the critical threshold, the entire curve shows smooth increases due to transient retrieval.

The transient-recovery curves reveal that going above capacity results in changes to retrieval performance that are far more graceful than equilibrium analyses might suggest. As $\alpha$ increases from below to above critical capacity, the curves change smoothly rather than exhibiting abrupt discontinuities (of course, however, the presence or absence of a flat plateau represents a qualitative difference between the two regimes). Thus, the ``blackout catastrophe'' is considerably less catastrophic when viewed through transient dynamics. Networks retain substantial memory function even beyond their critical capacities, provided one accepts transient rather than persistent recall.

While comparing individual curves at fixed $\alpha$ across different $n$ is not meaningful since $\alpha_c$ varies strongly with $n$, the families of curves generated by varying $\alpha$ can be compared. Comparing these families across interaction orders reveals important differences in transient retrieval characteristics. The Hopfield model exhibits the most graceful degradation, with curves that maintain roughly symmetric shape around the diagonal as $\alpha$ increases. In contrast, higher-order models show increasingly asymmetric behavior, with transient recovery effects becoming most pronounced for large $\bar{m}_{\text{init}}$. This asymmetry is most extreme for $n = 4$, where large initial overlaps are required to obtain substantial transient retrieval. Conversely, the Hopfield model shows significant transient recovery even for modest initial overlaps.

\subsection{Energy dynamics}
\label{subsec:energy_dynamics}

\begin{figure*}
    \centering
    \includegraphics[width=7in]{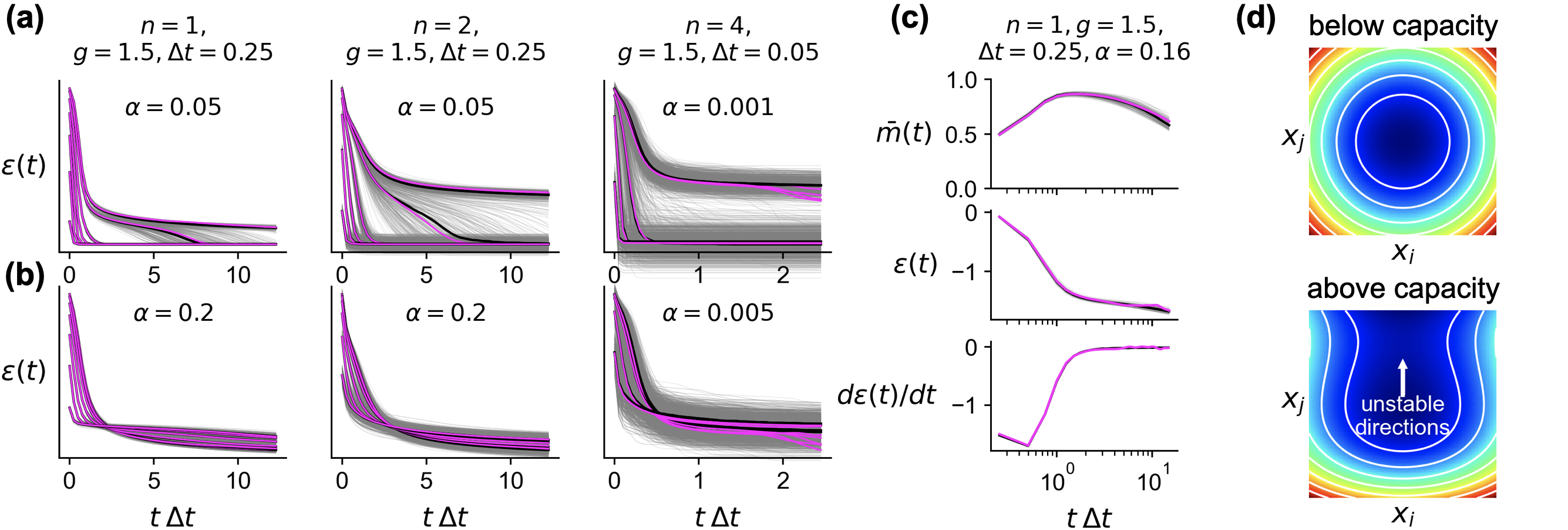}
\caption{Energy dynamics for different initial overlaps ($g = 1.5$). \textbf{(a)} Below-capacity dynamics with $\alpha < \alpha_c$. \textbf{(b)} Above-capacity dynamics with $\alpha > \alpha_c$. In (a) and (b), columns show $n=1, 2, 4$ from left to right. Different curves in each panel correspond to different initial normalized overlaps $\bar{m}_{\text{init}}$. {\color{gray}\textbf{Gray}} traces show individual finite-size simulations; \textbf{black} lines show simulation medians; and {\color{magenta}\textbf{magenta}} lines show DMFT predictions. \textbf{(c)} Example showing correspondence between energy decay and overlap evolution for $n=1$, $\alpha=0.16$. Top: normalized overlap $\bar{m}(t)$; middle: energy $\varepsilon(t)$; bottom: energy derivative $d\varepsilon(t)/dt$. Horizontal axis shows time on log scale. The fast rise and slow decay of the normalized overlap correspond to fast decay and slow decay of the energy, consistent with the system navigating shallow energy landscape features near stored patterns that eventually drive it away from the memory. \textbf{(d)} Schematic of the energy landscape structure near a stored pattern in the below-capacity regime, where a stable basin exists (top), and in the above-capacity regime, where a slow, unstable region persists despite the elimination of the stable basin (bottom).}
\label{fig:energy}
\end{figure*}

We now show that transient retrieval is caused by slow regions in the energy landscape near memories where, below capacity, stable fixed points previously existed.

\figref{fig:energy}{a,b} shows energy versus time for different initial overlaps, with columns representing $n = 1, 2, 4$ and (a) and (b) showing below- and above-capacity regimes, respectively. When entering stable retrieval states (below capacity with sufficient initial overlap), the energy quickly drops to a fixed value independent of the specific initial overlap. When displaying transient retrieval (above capacity or with small initial overlap), the energy exhibits two distinct timescales of decay: fast initial decay followed by slow decay. This slow decay corresponds to the system exploring regions of small gradient (that is, slow regions) of the energy landscape.

The correspondence between energy and overlap dynamics is illustrated in \figref{fig:energy}{c}, which shows three complementary views: normalized overlap $\bar{m}(t)$ (top), energy $\varepsilon(t)$ (middle), and energy time derivative $d\varepsilon(t)/dt$ (bottom), all shown on a logarithmic time scale. The fast energy decay corresponds to the fast rise of $\bar{m}(t)$, while the slow energy decay corresponds to the slow decay of $\bar{m}(t)$. This demonstrates that transient retrieval occurs when the system becomes trapped in slowly-varying regions of the energy landscape---residual features of the stable fixed points that existed below capacity. We schematize this landscape structure in \figref{fig:energy}{d}.

\subsection{Optimal readout time}
\label{subsec:optimal_readout}

\begin{figure}
    \centering
    \includegraphics[width=3.375in]{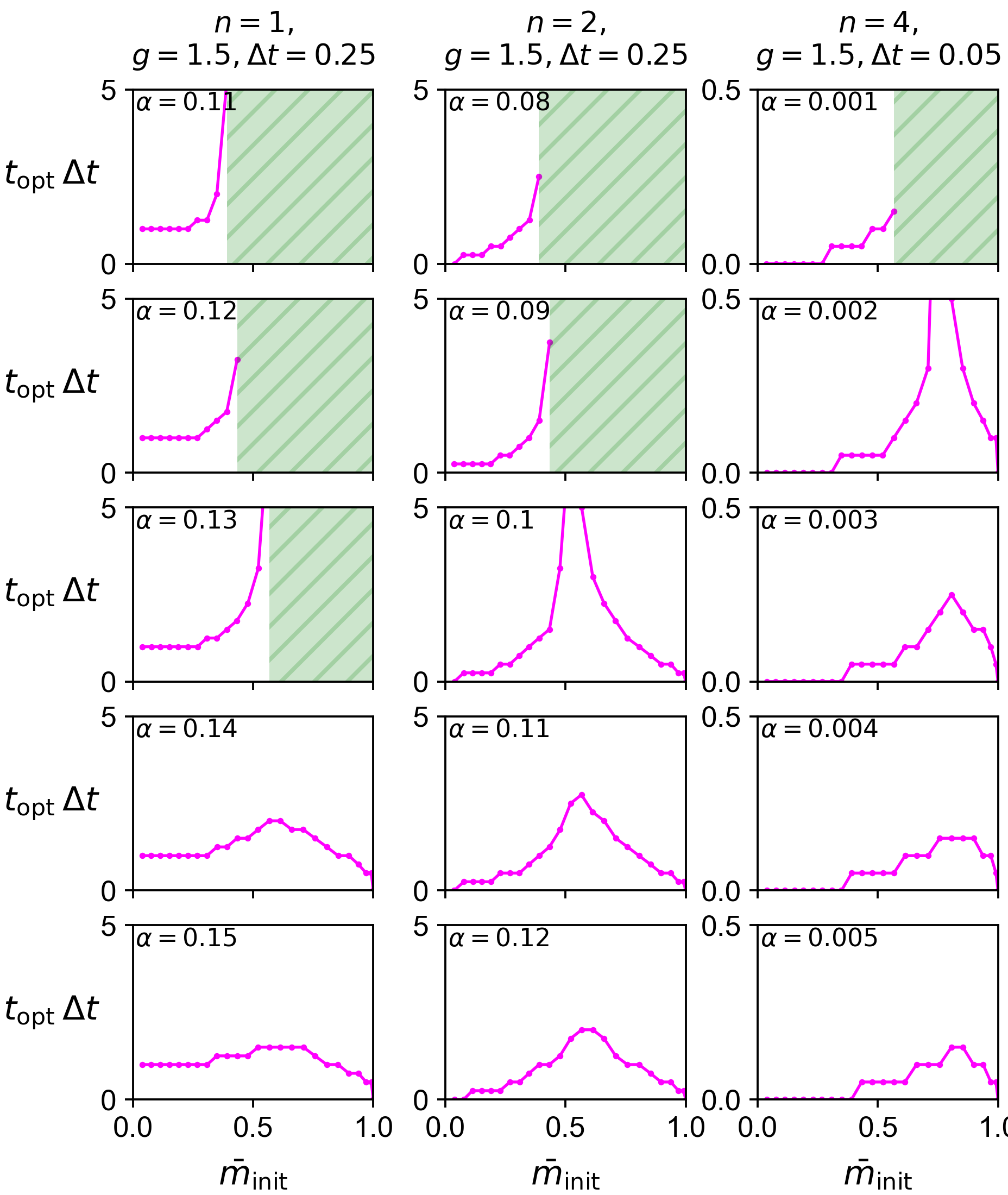}
    \caption{Optimal readout time $t_{\text{opt}} = \argmax_t \bar{m}(t)$ as a function of initial normalized overlap $\bar{m}_{\text{init}}$. Columns show $n=1, 2, 4$ from left to right. Rows show increasing values of $\alpha$. {\color{magenta}\textbf{Magenta}} lines show DMFT values of $t_{\text{opt}}$. {\color{ForestGreen}\textbf{Green}} shaded regions indicate where the network enters a stable retrieval state, making the optimal readout time undefined since any late time works equally well.}
\label{fig:how_long_to_wait}
\end{figure}

In networks storing memories as stable fixed points, memories can be read out at any sufficiently late time after convergence. However, when memories are retrieved via transient dynamics in slow regions of the energy landscape, there exists an optimal readout time that maximizes retrieval performance.

\figref{fig:how_long_to_wait}{} plots the optimal readout time,
\begin{equation}
    t_{\text{opt}} = \argmax_t \bar{m}(t),
\end{equation}
from the DMFT as a function of $\bar{m}_{\text{init}}$. Each row shows increasing values of $\alpha$, with columns showing $n = 1, 2, 4$. When the network enters a stable retrieval state for sufficiently high $\bar{m}_{\text{init}}$ below capacity, we shade the region and do not plot $t_{\text{opt}}$ since no well-defined unique optimal time exists---any sufficiently late time yields the same high overlap.

In networks below capacity (with shaded regions), $t_{\text{opt}}$ grows monotonically as $\bar{m}_{\text{init}}$ approaches the basin boundary. This growth becomes rapid and possibly divergent (though we cannot confirm this due to $T$ being finite in our solutions) as $\bar{m}_{\text{init}}$ approaches the critical value for entering the basin of attraction.

In networks above capacity (without shaded regions), no basin of attraction exists and the optimal readout time curves become non-monotonic. $t_{\text{opt}}$ increases for small $\bar{m}_{\text{init}}$, reaches a maximum at intermediate values---roughly where the below-capacity critical $\bar{m}_{\text{init}}$ was located---then decreases back toward $t_{\text{opt}} = 0$ at $\bar{m}_{\text{init}} = 1$. This non-monotonic behavior reflects the structure of the energy landscape: initial conditions with intermediate overlap can access the slow regions of the energy landscape near the stored pattern, making it worthwhile to wait as the system exploits these landscape features to increase overlap before eventual decay. In contrast, initial conditions with very high overlap are already near optimal alignment, while those with very low overlap cannot access the relevant landscape features. In these cases, there is little benefit to waiting, as the system either starts near its optimal retrieval performance or cannot meaningfully improve retrieval performance through transient dynamics.

\section{Discussion}
\label{sec:discussion}

\subsection{Historical methods}

While we have used the cavity method for its intuitive appeal and suitability to bipartite systems, most prior theoretical work on Hopfield dynamics has used path-integral approaches. We briefly review these methods, both to place our approach in context and to clarify the relationship between prior DMFT derivations and ours. One version of the path-integral approach, applicable to systems described by differential or difference equations (potentially with noise, e.g., Langevin dynamics), is known as the Martin--Siggia--Rose--De Dominicis--Janssen (MSRDJ) formalism \cite{martin1973statistical, de1978dynamics, hertz2016path}. This approach constructs a generating functional encoding correlation and response functions through its derivatives. The functional enforces the equations of motion using integral representations of delta functions. After averaging over the quenched disorder, order parameters are introduced that factor a resulting action across sites, and the order parameters are determined via saddle point at large $N$. This formalism has been used extensively to study the dynamics of disordered systems. For instance, \citet{sompolinsky1982relaxational} used it to study Langevin equations for a soft-spin version of the Sherrington-Kirkpatrick model, and it has found rich applications in disordered recurrent neural networks \cite{crisanti2018path, helias2020statistical, zou2024introduction}. The path-integral formalism also permits analysis of fluctuations around the $N\rightarrow \infty$ values of the order parameters, enabling computation of quantities such as the Lyapunov exponent \cite{crisanti2018path} or the dimension of activity \cite{clark2025connectivity} in random recurrent neural networks.

For discrete spins evolving at finite temperature, the natural description uses master equations (e.g., describing Glauber dynamics), and path-integral approaches have been developed for these systems as well. \citet{sommers1987path} developed one such method for the Sherrington-Kirkpatrick model, though this derivation was later questioned \cite{lusakowski1991comment}. Nevertheless, this formalism became the foundation for subsequent work on Hopfield dynamics. \citet{rieger1988glauber} used the Sommers formulation to study Hopfield dynamics but did not solve the DMFT equations, instead recovering replica-symmetric results of \cite{amit1985storing, amit1987statistical} through long-time limits. A similar approach was taken by \citet{horner1989transients}, who studied binary-spin Hopfield dynamics starting from a generating functional ``derived from Langevin dynamics of soft spin variables or from Sommers's formulation,'' then solved the DMFT equations using approximation schemes. For a review of these methods, see \citet{coolen1993dynamics} and more recently \citet{coolen2001statistical}. Another interesting work is that of \citet{gardner1987zero}, who used essentially the MSRDJ formalism to study deterministic, discrete-time evolution of binary spins in the Hopfield model (as well as the Sherrington-Kirkpatrick model and, briefly, higher-order generalizations of the Hopfield model; see below), deriving DMFT equations and analytically examining the first few time steps of retrieval. 

\subsection{Prior work on dynamics of associative memory models}

\subsubsection{Hopfield model}

As mentioned above, both Gardner and Horner had DMFT equations in hand, though their analyses had important limitations. \citet{gardner1987zero} derived full DMFT equations for the binary-spin Hopfield model with parallel dynamics but could only solve them analytically for the first few time steps, since the number of order parameters grows quadratically in time, rendering the analytics increasingly complicated.

Subsequent work sought to make progress by reducing the quadratic number of order parameters to a linear number at the cost of approximation. \citet{horner1989transients} started from the full two-time DMFT equations and derived an approximation scheme that reduces them to evolution equations for single-time quantities, interpolating between exact short-time behavior and the replica-symmetric stationary solution. The resulting equations could be integrated numerically to obtain the full overlap trajectory $m(t)$, but the dynamics at intermediate times are approximate. \citet{amari1988statistical} took a different approach, bypassing the DMFT entirely and instead appealing to heuristic Gaussianity assumptions about the noise experienced by each neuron. This yields evolution equations for a small number of macroscopic variables (the overlap and a variance), which also cannot capture the full two-time structure of the DMFT. Both works dealt with binary spins rather than the continuous variables studied here.

Both Horner and Gardner noted, but did not explore in detail, the nonmonotonic nature of retrieval that is central to our work. \citet{gardner1987zero} showed analytically that, for binary-spin Hopfield networks with parallel dynamics, the overlap with a stored pattern can increase during the first few time steps even above the critical capacity $\alpha_c \approx 0.14$, provided $\alpha < 2/\pi \approx 0.64$. \citet{horner1989transients} observed the same effect but described it as ``not very efficient'' and did not pursue it further. Finally, despite its heuristic nature, the work of \citet{amari1988statistical} is notable as an early, explicit analysis of transient retrieval phenomenology.

\subsubsection{Beyond Hopfield: dense models}

For dense associative memory models with higher-order interactions, \citet{gardner1987zero} provided the free energy as a function of order parameters to be determined via saddle point, and computed the overlap after the first time step. However, the authors did not write out the full DMFT equations for these models, in contrast to the standard Hopfield case for which they derived the complete DMFT.

During preparation of the present manuscript, \citet{mimura2026dynamical} provided a derivation, using path-integral methods, of DMFT equations for binary-spin dense associative memory models with deterministic dynamics. Their calculation should agree with the one initiated by \citet{gardner1987zero}. In contrast to our work, they studied binary-spin rather than graded-activity networks, used discrete time ($\dt = 1$) rather than continuous time, and derived the DMFT using generating functionals rather than the cavity method. They did not attempt to solve the equations numerically, instead appealing to an approximation that eliminates the self-coupling in the single-site dynamics, which is distinct from the approximation strategies of \citet{horner1989transients} and \citet{amari1988statistical} discussed above. They also did not analyze transient retrieval or energy dynamics, nor compare their results to simulations.

\subsubsection{Beyond Hopfield: vector models}

\citet{nicoletti2025statistical} studied, analytically and in simulations, a Hopfield model with $d$-dimensional vector spins (meaning that each spin is a vector confined to a $(d-1)$-sphere), focusing on how capacity and transient retrieval are affected by $d$. For $d=1$, one recovers binary spins, and the first-step retrieval formula of \citet{nicoletti2025statistical} recovers the corresponding result of \citet{gardner1987zero}, namely, that transient retrieval occurs for $\alpha < 2/\pi$. For $d>1$, \citet{nicoletti2025statistical} showed that the critical capacity for equilibrium retrieval \textit{shrinks} with $d$ ($\alpha_c \propto 1/d$), while the critical capacity for first-step transient denoising \textit{grows} with $d$ ($\tilde{\alpha} \propto d$). \citet{nicoletti2025statistical} note that they ``do not know why the first-step retrieval phase grows with spin dimension $d$'' and leave this as an open question, referencing the present paper's energy-landscape interpretation as a potential ``geometric memory hypothesis.'' We propose that increasing spin dimension enlarges the slow regions of the energy landscape near each pattern, which would simultaneously account for both the reduced static capacity (as enlarged slow regions of neighboring patterns interfere at lower loads) and the enhanced transient retrieval (as the slow regions become more prominent and persist to higher loads). Further probing this hypothesis is an interesting direction for future work.

\subsubsection{More analytically tractable associative memory models}

A rather different well-studied case of Hopfield model dynamics is that of randomly diluted networks, where a random mask is applied following Hebbian construction of the weights to zero out many matrix elements \cite{derrida1987exactly, derrida1987learning, tirozzi1991chaos, pereira2023forgetting}. In the limit where each neuron receives $K$ inputs with $K \rightarrow \infty$, $N \rightarrow \infty$, and $K/N \rightarrow 0$, the preactivations become Gaussian, allowing analytical solution similar to those for chaotic random recurrent networks \cite{sompolinsky1988chaos}. An interesting finding of \citet{pereira2023forgetting} is that, in this setting, overloading the network is associated with the emergence of chaotic attractors in the vicinity of stored patterns, a consequence of the random sparsity. In the fully connected models we study, overloading instead leads to the usual spin-glass phase governed by an energy function, and we characterize the dynamics on the resulting complex landscape.

One analytically tractable model of associative memory dynamics is that of \cite{bolle2003spherical}, in which all $N$ neuronal variables are jointly constrained to a sphere (as opposed to the vector model of \citet{nicoletti2025statistical}, where each individual neuron is a $d$-dimensional vector on its own sphere). Interactions are four-way in this model, since pairwise interactions do not allow a retrieval phase. In this case, one can use generating functional (or, presumably, cavity) techniques to derive differential equations governing the evolution of the two-time order parameters. This approach is similar to that used for $p$-spin glass models, whose DMFT and aging behavior were studied by \cite{altieri2020dynamical}.

\subsection{Dynamical perspective}

Equipped with the DMFT solutions derived and numerically solved in this work, we have taken a dynamical rather than an equilibrium perspective on associative memory models. While equilibrium analyses reveal the existence and stability of memory states, the dynamical view provides complementary insights into transient evolution, where much of the interesting memory retrieval actually occurs. We demonstrated that patterns can be transiently retrieved even when stable attractors no longer exist due to slow regions that persist in the energy landscape near stored patterns. The idea of using dynamics to probe energy landscape structure has precedent; Sompolinsky and colleagues \cite{sompolinsky1981time, sompolinsky1982relaxational} famously used finite-temperature dynamics of soft-spin glass models to uncover the ultrametric energy-landscape structure underlying the replica solution.

We introduced transient-recovery curves (\secref{subsec:transient_recovery}) as a tool for characterizing retrieval performance without requiring stable states, revealing that transient recall behavior changes gracefully as capacity increases rather than exhibiting an abrupt breakdown. Methods for characterizing or reverse-engineering recurrent neural networks whose computations rely on transient dynamics remain nascent \cite{turner2021charting} compared to fixed point-based methods \cite{maheswaranathan2019universality}, and this area warrants further attention.

The optimal time $t_{\text{opt}}$ before reading out a memory can be quite long, particularly when the system is initialized near the edge of a basin of attraction below capacity or near where this edge used to be above capacity (\secref{subsec:optimal_readout}). In this regime, multiple steps of recurrent dynamics contribute substantially to retrieval beyond the static information stored in the weights. Indeed, in very sparsely connected networks, multiple timesteps are required simply for information to propagate across the network \cite{turcu2022sparse}. A separate but related question is whether recurrent application of the same weights differs fundamentally from a deep feedforward architecture with untied weights; \citet{bauer2026unified} recently compared the two for learning general functions, though not retrieval specifically.

\subsection{Applications to other recurrent networks}

The analytical and numerical DMFT tools developed here should be useful for analyzing other large recurrent neural networks where transient dynamics play important roles. For example, minor modifications of our equations and numerical techniques allow for determining order parameters in randomly connected recurrent neural networks with varying levels of reciprocal correlation, $\rho = \langle J_{ij} J_{ji} \rangle / \langle J_{ij}^2 \rangle$. This problem was studied in simulations by \cite{marti2018correlations}, and the mean-field theory was solved under the assumption of stationary dynamics (dependence of order parameters only on $t-t'$) by \cite{clark2023dimension}. A similar approach was applied to the random Lotka-Volterra model in ecology by \cite{roy2019numerical}. For $\rho=1$, there exists a Lyapunov function and non-stationary behavior is guaranteed; such a system is analogous to zero-temperature dynamics in the Sherrington-Kirkpatrick model \cite{sompolinsky1982relaxational}. An open question remains whether non-stationary behavior persists (rather than eventually equilibrating to a time-translation invariant state) only at $\rho = 1$, for all $\rho$ above a critical $\rho_c > 0$, or for all $\rho > 0$.

\subsection{Exponential models}

The dense associative memory models we study achieve polynomial capacity, $P = \mathcal{O}(N^n)$. A separate line of work has studied associative memory models with exponential capacity, which arise when the energy function involves an exponential, rather than monomial, nonlinearity applied to the overlaps \cite{demircigil2017model, ramsauer2021hopfield, lucibello2024exponential}. These models have attracted significant interest due to their connection with the attention mechanism in transformers \cite{ramsauer2021hopfield}; for a recent review, see \cite{krotov2025modern}. The equilibrium properties of exponential-capacity models have been analyzed by \citet{lucibello2024exponential} using the observation that the noise term in the neuronal input takes the form of a random-energy-model free energy (see also \cite{zavatone2025nadaraya}). This analysis has been extended to structured patterns living on low-dimensional manifolds \cite{achilli2025capacity} (see also \secref{sec:struct_gen_creat}). As \citet{lucibello2024exponential} note, ``the full analytic computation of attraction basins requires following a trajectory in time. This is a complicated task, which has not been done in the standard Hopfield model, and which is beyond the reach of our method.'' Our DMFT framework addresses precisely this gap for both the Hopfield and higher-order polynomial-capacity cases, where the capacity scaling $P = \mathcal{O}(N^n)$ permits a controlled treatment of the noise from uncondensed patterns via expansions in $1/\sqrt{N}$. Extending the DMFT to exponential-capacity models is an interesting open problem; it would likely require treating a random-energy-model inner problem within the dynamical equations.

\subsection{From transient to stable retrieval}

An intriguing question is whether transient retrieval can be stabilized into persistent retrieval through additional mechanisms. \citet{gardner1987zero} suggested exploiting transient retrieval in \textit{below}-capacity networks to ultimately land the system in a basin of attraction: ``It might be possible after a few time steps of parallel iteration to define a way of annealing into a metastable state closer to the pattern.''

For \textit{above}-capacity networks, where no basin of attraction exists, a concrete realization of a biologically motivated stabilizing mechanism has recently appeared. In particular, \citet{del2025short} leveraged the theory of coupled neuronal-synaptic dynamics of \citet{clark2024theory}, in which Hebbian synaptic plasticity operates on a timescale comparable to neuronal dynamics (see also \cite{ba2016using}), to show that such ongoing Hebbian plasticity can deform the energy landscape during transient evolution, creating a stable minimum where previously only a slow, but unstable, region existed. The mechanism is analogous to a ball on a trampoline: the network's activity transiently visits the vicinity of a stored pattern, as our results predict for the static-weight case, and the fast plasticity simultaneously reshapes the local energy landscape to trap the system there.

More broadly, the interplay between multiple dynamical timescales in memory systems is an interesting direction \cite{wakhloo2025associative}, and the stabilization of transient retrieval through fast synaptic plasticity, as demonstrated by \citet{del2025short}, is one concrete example of what such interactions can achieve.

\subsection{Structured patterns, generalization, and creativity}
\label{sec:struct_gen_creat}

Most work on associative memory has aimed to retrieve only previously stored patterns, treating spurious states as failure modes. However, generalization requires the opposite, namely that the network develop \emph{new} attractors corresponding to patterns never seen during storage. Whether and how such generalization emerges depends on the structure of the stored data, and a growing body of equilibrium analyses has begun to map out the possibilities.

\citet{alemanno2023supervised} introduced a ``supervised'' Hebbian learning rule in which the synaptic matrix is constructed from class-mean patterns rather than individual examples, establishing an equivalence of the resulting system with restricted Boltzmann machines. When training examples are noisy versions of underlying centroids, a generalization phase emerges in which individual noisy examples do not form attractors but the centroids do. \citet{agliari2024spectral} considered a related setting using random-matrix theory, studying the eigenvalue distribution of the weight matrix built from noisy examples, relating spectral properties to retrieval and generalization. \citet{mezard2017mean} analyzed Hopfield networks storing patterns that reside on a low-dimensional subspace using a bipartite cavity method, where generalization corresponds to attractors spanning the subspace. More recently, \citet{kalaj2025random} studied a related model with richer structure, the random-features Hopfield model, where patterns are generated through a nonlinear projection of latent variables. In the above-capacity regime, the network develops attractors corresponding to previously unseen combinations of learned features, namely, uniform combinations of subsets of features. These would classically be considered spurious states but here correspond to meaningful points on the latent manifold. Related behavior has been observed in the context of diffusion models \cite{pham2025memorization, kadkhodaie2024generalization, kamb2025analytic}. For dense associative memory models with exponential nonlinearities, \citet{achilli2025capacity} have studied capacity under a conceptually similar data manifold hypothesis \cite{goldt2020modeling}; for patterns on a linear manifold, they showed analytically that the entire manifold becomes the sole attractor. In each of these settings, what would classically be considered a spurious state is instead interpreted as a meaningful generalization of the training data.

A separate line of work has modified the learning rule itself to improve capacity or promote generalization. \citet{fachechi2019dreaming, agliari2019dreaming} introduced a ``dreaming'' construction for Hopfield network weights, defining a family of interaction matrices parameterized by a ``dreaming time'' $t_d$ that interpolates between the Hebbian rule ($t_d = 0$) and the pseudo-inverse rule ($t_d \to \infty$), increasing the storage capacity from $\alpha_c \approx 0.14$ to $\alpha_c \approx 1$. \citet{agliari2024regularization} showed that gradient descent on a regularized loss for Hopfield networks yields this same family of interaction matrices, with $t_d$ equal to the inverse of the regularization hyperparameter. This optimization perspective reveals that moderate regularization (intermediate $t_d$) enables generalization, with attractors forming at ground-truth centroids rather than individual noisy examples (similar to the supervised Hebbian setting of \citet{alemanno2023supervised}), while excessive regularization (large $t_d$) causes overfitting. \citet{serricchio2025daydreaming} introduced a distinct iterative algorithm called ``daydreaming'' that simultaneously reinforces stored patterns and erases spurious states at each learning step. On correlated data, it exploits correlations to increase storage capacity and basin sizes. Finally, \citet{d2025pseudo} showed that maximizing a pseudo-likelihood at zero temperature naturally produces an associative memory that, for structured datasets, transitions from memorization to generalization as training examples increase, developing attractors for unseen data. In a related setting, \citet{d2024self} trained a recurrent self-attention network via pseudo-likelihood and observed that both training and test examples appear as transient states of the dynamics.

All of the analyses described above are equilibrium or static in nature. Extending the dynamical analysis of the present paper to these structured and learned settings is a natural open direction. On structured data, the transient dynamics studied here may take on new significance: rather than transiently visiting the vicinity of a stored pattern before being driven away, the system might transiently explore ``generalization regions'' (centroids, feature mixtures, or latent manifolds) that reflect the underlying data structure. The methods developed here could reveal how networks navigate structured energy landscapes during retrieval, how dynamics interact with hierarchical or compositional structure in stored data, and whether transient retrieval properties are qualitatively different when the landscape has been shaped by learning rules that promote generalization.

More speculatively, one might view creative thought as a dynamical process on a complex energy landscape shaped by past experience, where the goal is not to converge to a minimum corresponding to a previously stored memory, but to explore the landscape's structure in novel ways. 

\subsection{Outlook}

We have considered noiseless dynamics throughout this work. Understanding how noise affects transient retrieval is an interesting direction. \cite{behera2023enhanced} showed that temporal correlations in noise can improve retrieval in the higher-order spherical generalization of the Hopfield model of \cite{bolle2003spherical}.

Finally, neuroscience experiments could attempt to test whether biological neural networks exploit transient dynamics for memory retrieval, as our results suggest they could. Experimental signatures of this phenomenon could be detected through analysis of neural population activity during memory tasks. Concretely, one could compute time-varying similarity measures between population activity and stored memory patterns, analogous to our overlap function $m(t)$. The presence of transient retrieval would manifest as initial increases in pattern similarity followed by slower decay. Observing this on error trials would be a particularly compelling connection to behavior.

\section{Acknowledgments}

The author thanks Jacob Zavatone-Veth, Blake Bordelon, L.F. Abbott, Ken Miller, Stefano Fusi, Ashok Litwin-Kumar, and Haim Sompolinsky for valuable discussions, comments, suggestions, and pointers. The author was supported during preparation of this work by the Gatsby Charitable Foundation and the Kavli Foundation.

\vspace{0.3cm}

\noindent\textbf{Code} for reproducing the results in this paper is available at \url{https://github.com/davidclark1/TransientDynamicsAssocMem}.

\appendix

\section{Numerically stable \texorpdfstring{$\mathcal{F}(\phi)$}{F(phi)}}
\label{app:comp_F}

For $\phi(x) = \tanh(x)$, $\mathcal{F}(\phi)$ can be evaluated in a numerically stable manner using
\begin{align}
    \mathcal{F}(\phi) &= \frac{1}{2}\log\left(1 - \phi^2\right) + x \phi \\
    &= \log 2 + x - \text{SoftPlus}(2x) + x\phi,
\end{align}
where $\tanh x = \phi$ and $\text{SoftPlus}(x) = \log(1 + e^{x})$ has numerically stable implementations in common libraries.

\section{Fixed-point mean-field theory and critical capacities}
\label{app:equilibrium}

For completeness, we derive the fixed-point mean-field theory that yields the critical capacities reported in the main text. Equilibrium properties of the Hopfield model were computed using the replica method by \citet{kuhn1991statistical}, and our results for $n=1$ should agree with this analysis. Here we obtain the fixed-point statistics for general $n$ by taking the static limit of our DMFT cavity equations. This analysis confirms that the model reproduces the ``blackout catastrophe''---the discontinuous vanishing of the nonzero overlap solution.

For a single condensed pattern, the equilibrium single-site equations result from removing time dependence from the DMFT equations, yielding the self-consistent system
\begin{align}
    x &= \frac{g}{\sqrt{\alpha}}\xi m^n + \eta + F \phi(x), \label{eq:eq_sst} \\
    m &= \tavg{\xi \phi(x)}_\sst, \label{eq:eq_overlap} \\
    C^\phi &= \tavg{\phi^2(x)}_\sst, \label{eq:eq_correlation} \\
    S^\phi &= \tavg{\frac{\phi'(x)}{1 - F \phi'(x)}}_{\sst}, \label{eq:eq_response}
\end{align}
where we have dropped the subscript $0$ for brevity and set $\sigma_\xi = 1$. The cavity field variance and self-coupling kernel are given by
\begin{align}
    {C}^{\eta} &= \begin{cases}
        g^2 K^{2} C^\phi & n=1 \\
         (2n-1)!! \: g^2 (C^\phi)^n  & n > 1 \text{, even}
    \end{cases} \label{eq:eq_C_eta} \\
    {F} &= \begin{cases}
        g \sqrt{\alpha} K &  n=1 \\
         n \:(2n-1)!! \: g^2 S^{\phi} (C^\phi)^{n-1}   &  n > 1 \text{, even}
    \end{cases} \label{eq:eq_F}
\end{align}
where for the Hopfield case ($n=1$), we have defined
\begin{equation}
    K = \frac{1}{1 - \frac{g}{\sqrt{\alpha}} S^\phi}.
\end{equation}
The single-site average $\tavg{\cdot}_\sst$ involves sampling $\xi \sim P(\xi)$ and $\eta \sim \mathcal{N}(0, C^\eta)$, then solving for $x$ according to the nonlinear equation \eref{eq:eq_sst}.

Using $\phi(x) = \tanh(x)$, an issue arises when $F \geq 1$, as the equation $x = F\phi(x) + A$ (where $A = \frac{g}{\sqrt{\alpha}}\xi m^n + \eta$) can have multiple solutions for a range of $A$ values. This issue is resolved in the replica approach by requiring that the chosen solution minimizes a certain energy function \cite{kuhn1991statistical}. There should be a way to recover this prescription in the cavity approach, making our equations agree with those of \citet{kuhn1991statistical} for $n=1$ in our theory and temperature $\rightarrow 0$ in theirs, modulo self-interactions, but it is not immediately obvious how to do this.

To sidestep this multiple-solution issue, we restrict our analysis to $g = 1.5$, for which we find that $F$ remains substantially below unity until the overlap solution $m$ abruptly vanishes with increasing $\alpha$. This allows us to compute the critical capacities where the high-overlap solution disappears (Table~\ref{tab:critical_capacities}).
\begin{table}[ht!]
\centering
\begin{tabular}{|c|c|}
\hline
Interaction order $n$ & Critical capacity $\alpha_c$ (at $g = 1.5$) \\
\hline
1 & 0.13 \\
2 & 0.080 \\
4 & 0.0011 \\
\hline
\end{tabular}
\caption{Critical capacities at $g = 1.5$.}
\label{tab:critical_capacities}
\end{table}

\section{Effects of \texorpdfstring{$g$}{g} and \texorpdfstring{$\dt$}{dt}}
\label{sec:gain}

\begin{figure}
    \centering
    \includegraphics[width=3.375in]{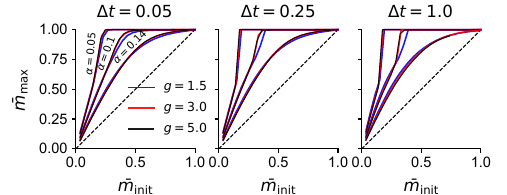}
    \caption{Effects of gain $g$ and time step $\dt$ on transient retrieval dynamics for the Hopfield model ($n=1$). Three panels from left to right show $\dt = 0.05, 0.25, 1.0$. Transient-recovery curves are shown for $\alpha = 0.05, 0.1, 0.14$, with gain values $g = 1.5, 3, 5$ distinguished by different colors for each $\alpha$. Axes show maximum normalized overlap $\bar{m}_{\text{max}}$ versus initial normalized overlap $\bar{m}_{\text{init}}$ as in \figref{fig:trans_recov_curves}{}. Curves for different gain values at the same $\alpha$ are closely grouped.}
    \label{fig:g_behavior}
\end{figure}

The transient retrieval phenomenology depends on the model parameters $g$ and $\dt$. \figref{fig:g_behavior}{} explores how varying these parameters affects transient retrieval through transient-recovery curves for the Hopfield model. Once $g$ reaches $\sim 3$, the curves largely converge. No clear trends are visible regarding $\dt$.

\section{Odd \texorpdfstring{$n$}{n} and scaling behavior}
\label{sec:odd_n}

\begin{figure}
    \centering
    \includegraphics[width=\columnwidth]{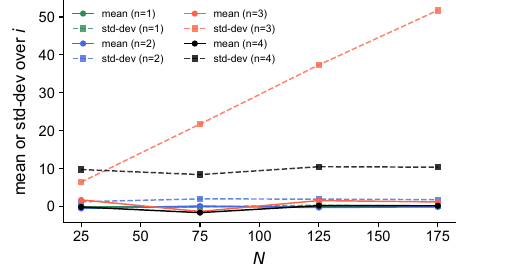}
    \caption{Demonstration of divergent behavior for odd interaction orders. Mean and standard deviation of neuronal input $\text{input}_i = \frac{1}{\sqrt{\alpha}}\sum_\mu \xi^\mu_i (m^\mu)^{n}$ as functions of system size $N$ for interaction orders $n = 1, 2, 3, 4$.}
    \label{fig:odd_n}
\end{figure}

To demonstrate the divergent behavior that occurs with odd $n$, we compute synthetic neuronal inputs, $\text{input}_i = \frac{1}{\sqrt{\alpha}}\sum_\mu \xi^\mu_i (m^\mu)^{n}$, where $m^\mu = \frac{1}{N}\sum_i \xi^\mu_i \phi_i$ and both $\xi^\mu_i$ and $\phi_i$ are iid $\pm 1$. We use $\alpha = 0.01$ and $P = \alpha N^n$. We compute this for each combination of $N \in \{25, 75, 125, 175\}$ and $n \in \{1, 2, 3, 4\}$. The mean and standard deviation of the neuronal input across the $i$ index for a single draw of all variables are plotted in \figref{fig:odd_n}{}. With increasing $N$, the standard deviations remain flat (consistent with proper scaling) for all cases except $n = 3$, which grows linearly with $N$.

\bibliography{refs}

\end{document}